\documentclass[12pt]{article}
\usepackage{jheppub}
\usepackage{mathtools,amssymb,amsthm}
\usepackage[utf8]{inputenc}
\usepackage{enumitem}
\usepackage{xcolor}
\usepackage{graphicx}
\usepackage{multirow}
\usepackage[normalem]{ulem}
\usepackage{longtable}
\usepackage{array}
\allowdisplaybreaks

\usepackage{tikz}
\usetikzlibrary{decorations.pathmorphing}

\usepackage{overpic}

\newcommand{\be}{\begin{equation}}
\newcommand{\ee}{\end{equation}}
\newcommand{\bi}{\begin{itemize}}
\def \bea {\begin{eqnarray}}
\def \eea {\end{eqnarray}}

\def\ba#1\ea{\begin{align}#1\end{align}}
\def\bad#1\ead{\begin{aligned}#1\end{aligned}}
\def\bg#1\eg{\begin{gather}#1\end{gather}}
\def\bm#1\em{\begin{multline}#1\end{multline}}
\def\bmd#1\emd{\begin{multlined}#1\end{multlined}}

\usepackage{color}

\newcommand{\ignore}[1]{}

\def\d{\delta}

\def \bal#1\eal  {\begin{align} #1 \end{align}}
\def \bga#1\ega  {\begin{gather} #1 \end{gather}}
\def\({\left(}
\def\){\right)}
\def\[{\left[}
\def\]{\right]}
\def\<{\left\langle}
\def\>{\right\rangle}
\def\d{\mathrm{d}}

\newcommand{\eim}{\end{itemize}}
\newcommand{\beq} {\begin{equation}}
\newcommand{\eeq} {\end{equation}}
\newcommand{\bc}{\begin{center}}
\newcommand{\ec}{\end{center}}

\begin{document}

\title{Quantum corrections to the path integral of near extremal de Sitter black holes}

\author[a]{Matthew J. Blacker}\author[a]{, Alejandra Castro}\author[a,b]{, Watse Sybesma}\author[c,d,e]{ and Chiara Toldo}
\affiliation[a]{Department of Applied Mathematics and Theoretical Physics, University of Cambridge, Cambridge CB3 0WA, United Kingdom}
\affiliation[b]{Nordita, KTH Royal Institute of Technology and Stockholm University,\\ Hannes Alfvéns väg 12, SE-106 91 Stockholm, Sweden}
\emailAdd{mjb318@cam.ac.uk, ac2553@cam.ac.uk, watse.sybesma@su.se, chiaratoldo@fas.harvard.edu}
\affiliation[c]{Department of Physics, Harvard University, 17 Oxford Street, Cambridge MA 02138, USA}
\affiliation[d]{Dipartimento di Fisica, Universita' di Milano, via Celoria 6, 20133 Milano MI, Italy}
\affiliation[e]{INFN, Sezione di Milano, Via Celoria 16, I-20133 Milano, Italy}

\abstract{
We study quantum corrections to the Euclidean path integral of charged and static four-dimensional de Sitter (dS$_4$) black holes near extremality. These black holes admit three different extremal limits (Cold, Nariai and Ultracold) which exhibit AdS$_2 \times S^2 $, dS$_2 \times S^2 $  and $\text{Mink}_2 \times S^2$ near horizon geometries, respectively. The one-loop correction to the gravitational path integral in the near horizon geometry is plagued by infrared divergencies due to the presence of tensor, vector and gauge zero modes. Inspired by the analysis of black holes in flat space, we regulate these divergences by introducing a small temperature correction in the Cold and Nariai background geometries. In the Cold case, we find a contribution from the gauge modes which is not present in previous work in asymptotically flat spacetimes. Several issues concerning the Nariai case, including the presence of negative norm states and negative eigenvalues, are discussed, together with problems faced when trying to apply this procedure to the Ultracold solution.

}

\maketitle

%
\section{Introduction}
%
The Euclidean path integral approach to quantum gravity, pioneered by
\cite{gibbons_action_1977}, provides persuasive rules to quantify effects in quantum gravity. 
 The basic idea is to take a semi-classical approach, where one evaluates the path integral by a saddle point approximation;  this has arguably provided insights into how gravity interplays with the quantum arena. In the context of black holes, this procedure has been powerful since there are quantum effects that can be quantified by the low-energy effective theory without requiring a UV completion.  One renowned example is the analysis developed for supersymmetric black holes in \cite{Banerjee:2010qc}, and applied more broadly in, e.g., \cite{Sen:2012cj,Sen:2012dw}.

More recently, it has been demonstrated that the path integral of certain perfectly smooth Euclidean black holes at low temperature is dominated by quantum effects  \cite{Charles:2019tiu,Iliesiu:2020qvm,Heydeman:2020hhw,Iliesiu:2022onk}. Concretely it was noticed that the AdS$_2$ geometry, which usually describes a portion of the geometry of black holes at zero temperature, is plagued by zero modes which introduce an infrared divergence in the path integral. Regulating this divergence by heating up the black hole shows that these zero modes are corrected.
Moreover, these quantum corrections surpass the effects of the classical action by adding (in the canonical ensemble) a log-$T$ term in the quantum effective action, where $T$ denotes the temperature of the black hole. This dramatically changes the density of states, and places in a new light the quantum nature of black holes at zero temperature.\footnote{ Describing the microscopic properties of black holes at zero temperature is challenging and potentially confusing, which goes back to the work of \cite{Preskill:1991tb,Maldacena:1998uz,Page:2000dk}. See \cite{Turiaci:2023wrh} for a recent discussion on this, which covers how these recent developments resolve those old challenges.}

Zero temperature, or to be more precise, zero surface gravity at a horizon, does not necessarily always lead to an AdS$_2$ factor in the black hole geometry. When black holes are embedded in a theory with a positive cosmological constant, other possibilities can arise. Our aim here is to quantify the behaviour of the Euclidean path integral in such circumstances and establish the robustness of the log-$T$ effects more broadly.   

In this paper, we consider a Reissner-Nordström-de-Sitter black hole (RNdS$_{4}$), which is a static solution of Einstein-Maxwell theory with a positive cosmological constant.  These black holes have a region of parameter space where three horizons can be at play (in decreasing order): a cosmological horizon $r_c$, an outer horizon $r_+$, and an inner horizon $r_-$. This allows for the definition of three different zero temperature limits, which define an extremal black hole. The \textit{Cold} black hole corresponds to $r_+=r_-$, and there is indeed a portion of the geometry described by AdS$_2$ near the extremal horizon. The case of $r_c=r_+$ will be coined \textit{Nariai}, and the near horizon geometry has a dS$_2$ factor. Finally, we have the \textit{Ultracold} black hole for which $r_c=r_+=r_-$ and one encounters a factor of Mink$_2$ in the near horizon region. As a special case, we will also consider Schwarzschild-de Sitter (SdS$_{4}$), for which $r_-$ is not present. The extremal limit of SdS$_{4}$ is a Nariai geometry.\footnote{\cite{Guerrero-Dominguez:2022riz} also computes the logarithmic corrections to Schwarzschild-de Sitter partition function. There, however, the path integral is perturbed by the addition of a scalar field on the near-extremal background geometry, as opposed to the graviton excitations which we consider in Sec.\,\ref{SecSdSNariai}.} This gives us different configurations where we would like to quantify how (or if) the log-$T$ appears in the gravitational path integral. 

In this work we will focus on the quantum correction of the path integral at low temperatures for the Cold and Nariai black holes, and focus on the near-horizon geometry of these solutions where the AdS$_2$ and dS$_2$ factors are manifest. We will make some comments in our discussion about the Ultracold black hole since it is not a simple case to analyse. We will be adopting the method proposed in \cite{Iliesiu:2022onk} together with \cite{Banerjee:2023quv}, which goes as follows. The first task is to establish that there is an infrared divergence at $T=0$ in the Euclidean path integral. This arises by looking at one-loop corrections to the Einstein-Maxwell theory in the near-horizon region. A divergence will occur if we encounter an infinite number of eigenstates for the operators appearing at one-loop in the effective action that carry a zero eigenvalue. These encompass the so-called ``zero-modes.''  The method to regulate this divergence is to turn on a deformation, usually by heating the black hole, and this corrects the eigenvalues. 
Schematically, after the correction we would have that the Euclidean path integral is
\begin{equation}\label{eq:intropartition}
    Z_{\text{low-}T}
    \sim
    \int 
    \mathcal{D}h_{\text{z.m.}}
    \mathcal{D}a_{\text{z.m.}}
    e^{
        -\delta\Lambda_{h}(T)\langle h_{\text{z.m.}}|h_{\text{z.m.}}\rangle
        -\delta\Lambda_{a}(T)\langle a_{\text{z.m.}}|a_{\text{z.m.}}\rangle
    } ~,
\end{equation}
where we have denoted with $h_{\text{z.m.}}$ the graviton zero modes, and with $a_{\text{z.m.}}$ the gauge zero modes. The expressions $\delta \Lambda_{h,a}(T)$ denote the first-order correction in temperature $T$ of the eigenvalue of the related mode. Up to one-loop, there is also a tree-level contribution and contributions from modes with non-zero eigenvalue; we will not report these contributions to the path integral. Our aim here is to establish that there is a strong quantum effect at low-$T$ for RNdS$_4$ black holes and potential obstacles that differ from the standard cases studied in the literature so far.  

The outcome of our treatment is summarized in Table \ref{fig:table1}, where one can see that the Cold case exhibits consistently positive norm zero modes, which also acquire always a positive eigenvalue correction upon switching on temperature. In this case the path integral is convergent and we can quantify clearly the log-$T$ effect. The Nariai solution is instead plagued with pathologies: first of all, we find that gauge modes, when present, have always negative norms. Tensor modes, in the canonical Wick-rotated limit of dS$_2 \times S^2$ geometry, have negative eigenvalue corrections, and vector modes have negative norms. One way to circumvent this problem is to work with a complexified near horizon limiting geometry, however some issues persist due to negative norm gauge modes. We refer the reader to the respective sections for a detailed treatment.
Our work in the Nariai case should be contrasted with \cite{Maldacena:2019cbz,Cotler:2019nbi,Moitra:2022glw}, where these corrections are reported for Jackiw-Teitelboim (JT) gravity \cite{Teitelboim:1983ux,Jackiw:1984je} with a positive cosmological constant, which is a two-dimensional theory containing dS$_2$ as a solution. JT gravity only captures the tensor mode contributions reported here. The recent work in \cite{Turiaci:2025xwi} is an appropriate point of comparison for the case of SdS$_4$.

\begin{table}[ht]
\begin{center}
\begin{tabular}{|c|cc|cc|cc|}
\hline
\multirow{2}{*}{Decoupling limits}                                               & \multicolumn{2}{c|}{Tensor modes}  & \multicolumn{2}{c|}{Vector modes}  & \multicolumn{2}{c|}{Gauge modes}   \\ \cline{2-7} 
                                                                                 & \multicolumn{1}{c|}{$\langle h_{\text{z.m.}}|h_{\text{z.m.}}\rangle$} & $\delta\Lambda_{h}(T)$ & \multicolumn{1}{c|}{$\langle h_{\text{z.m.}}|h_{\text{z.m.}}\rangle$} & $\delta\Lambda_{h}(T)$ & \multicolumn{1}{c|}{$\langle a_{\text{z.m.}}|a_{\text{z.m.}}\rangle$} & $\delta\Lambda_{a}(T)$  \\ \hline \hline
\begin{tabular}[c]{@{}c@{}}RNdS$_4$ Cold\\ Sec. \ref{SecColdLimit} \end{tabular}                       & \multicolumn{1}{c|}{+}     & +     & \multicolumn{1}{c|}{+}     & +     & \multicolumn{1}{c|}{+}     & +     \\ \hline
\begin{tabular}[c]{@{}c@{}}SdS$_4$ Nariai\\ Sec.  \ref{SecSdSNariai}\end{tabular}                     & \multicolumn{1}{c|}{+}     & $-$     & \multicolumn{1}{c|}{$-$}     & $-$     & \multicolumn{1}{c|}{}      &       \\ \hline
\begin{tabular}[c]{@{}c@{}}SdS$_4$ Nariai\\ (Complexified)\\ Sec. \ref{SubSecWickRot}\end{tabular} & \multicolumn{1}{c|}{+}     & +     & \multicolumn{1}{c|}{$-$}     & +     & \multicolumn{1}{c|}{}     &      \\ \hline
\begin{tabular}[c]{@{}c@{}}RNdS$_4$ Nariai\\ Sec. \ref{sec:decoup61}\end{tabular}                  & \multicolumn{1}{c|}{+}     & $-$     & \multicolumn{1}{c|}{$-$}     & $\pm$    & \multicolumn{1}{c|}{$-$}     & $+$     \\ \hline
\begin{tabular}[c]{@{}c@{}}RNdS$_4$ Nariai\\ (Complexified)\\ Sec. \ref{SubSecWickRot}\end{tabular} & \multicolumn{1}{c|}{+}     & +     & \multicolumn{1}{c|}{+}     & +     & \multicolumn{1}{c|}{$-$}     &  $-$    \\ \hline
\end{tabular}
\end{center}
\caption{
In this table we present the signs of different contributions to the partition function \eqref{eq:intropartition} for different decoupling limits of Reissner-Nordström-de Sitter. The results for a near-extremal asymptotically AdS$_{4}$ Reissner-Nordström black hole in Sec.\,\ref{AdS4_subs} can be related to the Cold black hole by flipping the sign of the four dimensional cosmological constant.
The $\pm$ sign for the vector modes depends on whether the black hole resides below or above the lukewarm curve in the phase diagram, see Fig.\,\ref{fig:sharkfin}.}
\label{fig:table1}
\end{table}

The paper is organized as follows. We start in Sec.\,\ref{sec:dSRN}  with the description of static de Sitter black holes in four dimensions, discussing their thermodynamic quantities, extremal limits (Cold, Nariai and Ultracold) and near horizon geometries.  In Sec.\,\ref{Sec3GenStrat} we explain our setup and we describe the saddle point approximation and the quantum corrections to the path integral in the near horizon geometry. We spell out the constraints obeyed by the zero modes and their physical interpretation, together with the regularization strategy, which employs a near-extremal geometry as a regulator. In Sec.\,\ref{SecColdLimit} we then treat the log-$T$ corrections to the near horizon path integral in the Cold geometry, including those due to tensor zero modes, vector zero modes and the gauge zero modes. We also include in that section the corrections for the extremal AdS$_4$ black hole case, for comparison to dS$_4$. Sec.\,\ref{SecSdSNariai} is then devoted to the study of the quantum corrections in the Nariai limit of the Schwarzschild-de Sitter black hole. The corrections to the path integral in the Nariai limit of RNdS$_4$ are treated in Sec.\,\ref{SecNariaiCorrections}, where we treat two different cases: the first one corresponds to the canonical Euclidean Wick rotation of the dS$_2 \times S^2$ near horizon geometry, while the second is a complexified geometry that still solves the Einstein's equations with charge and positive cosmological constant, but differs from the EAdS$_2 \times S^2$ background by an overall minus sign.  We conclude with an outlook section, pointing out the salient findings of our paper and future directions.

\paragraph{Note added:}   During the write-up of our work, the preprint \cite{Maulik:2025phe} appeared on the arXiv. There is some overlap with our analysis on the computation of the corrections due to the tensor zero modes for Cold and Nariai near-extremal geometries. That work does not consider the vector or gauge modes which play a significant role in our discussion.

\section{Static \texorpdfstring{dS$_{4}$}{dS4} black holes}\label{sec:dSRN}

In this section, we review some basic properties of four-dimensional static black holes in Einstein-Maxwell theory with a positive cosmological constant that will be used subsequently. In the first portion we describe the non-extremal black holes, which are the solutions described in \cite{Romans:1991nq,Mann:1995vb,Booth:1998gf}.  In the remainder of this section, we provide details of the various extremal cases, and their near-horizon geometries, which will be the relevant backgrounds for the computation of the quantum corrections described in this paper. This section follows the conventions of \cite{Castro:2022cuo}.

\subsection{Einstein-Maxwell theory with a positive cosmological constant}
\label{SecRNdSandSdS}
The class of (near-)extremal black holes we will study arise as solutions to the equations of motion of Einstein-Maxwell theory in four dimensions with a positive cosmological constant. The action is given by
\begin{equation}\label{eq:vanilla4daction}
    S_{\text{EM}}=\frac{1}{16\pi G_{N}}\int \d^{4}x\,\sqrt{-g}
    \Big(R-2\Lambda-F_{\mu\nu}F^{\mu\nu}\Big)\,,
\end{equation}
where $R$ is the Ricci scalar, $\Lambda=3/\ell^{2}_{4}$ is the cosmological constant and $F_{\mu\nu}$ is the field strength of gauge field $A_{\mu}$. Throughout this work, we will use units $G_N = 1$. 

\begin{figure}[ht]
\vspace{0.3cm}
\begin{center}
\begin{tikzpicture}[scale=0.7]

	\path [fill=lightgray] (0,0) -- (-4,4)  -- (4,4);
	\path [fill=lightgray] (0,0) -- (4,-4)  -- (-4,-4);
	
	\path [fill=lightgray] (8,0) -- (4,4) -- (8,8) -- (12,4);
	\path [fill=lightgray] (8,0) -- (4,-4)  -- (12,-4);
	
	\path [fill=lightgray] (8,8) -- (4,12)  -- (12,12);

	\draw (4,12) circle [radius =0.05];
	\draw (4.5,12) circle [radius =0.05];
	\draw (5.,12) circle [radius =0.05];
	\draw (5.5,12) circle [radius =0.05];
	\draw (6,12) circle [radius =0.05];
	\draw (6.5,12) circle [radius =0.05];
	\draw (7,12) circle [radius =0.05];
	\draw (7.5,12) circle [radius =0.05];
	\draw (8,12) circle [radius =0.05];
	\draw (8.5,12) circle [radius =0.05];	
	\draw (9,12) circle [radius =0.05];
	\draw (9.5,12) circle [radius =0.05];	
	\draw (10,12) circle [radius =0.05];
	\draw (10.5,12) circle [radius =0.05];
	\draw (11,12) circle [radius =0.05];
	\draw (11.5,12) circle [radius =0.05];
	\draw (12,12) circle [radius =0.05];
	
	\draw (4,-4) circle [radius =0.05];
	\draw (4.5,-4) circle [radius =0.05];
	\draw (5.,-4) circle [radius =0.05];
	\draw (5.5,-4) circle [radius =0.05];
	\draw (6,-4) circle [radius =0.05];
	\draw (6.5,-4) circle [radius =0.05];
	\draw (7,-4) circle [radius =0.05];
	\draw (7.5,-4) circle [radius =0.05];
	\draw (8,-4) circle [radius =0.05];
	\draw (8.5,-4) circle [radius =0.05];	
	\draw (9,-4) circle [radius =0.05];
	\draw (9.5,-4) circle [radius =0.05];	
	\draw (10,-4) circle [radius =0.05];
	\draw (10.5,-4) circle [radius =0.05];
	\draw (11,-4) circle [radius =0.05];
	\draw (11.5,-4) circle [radius =0.05];
	\draw (12,-4) circle [radius =0.05];

	\draw[decorate,decoration=zigzag] (4,4)--(4,12);
	\draw[decorate,decoration=zigzag] (12,4)--(12,12);
	\draw (4,4)--(12,12); 
	\draw (12,4)--(4,12); 
	
	\draw (0,0)--(4,4);
	\draw (0,0)--(4,-4);
	\draw (0,0)--(-4,-4);
	\draw (0,0)--(-4,4);
	
	\draw (4,4)--(12,-4); 
	\draw (4,-4)--(12,4); 

	\draw (-4,4)--(4,4);	
	
	\draw (-4,-4)--(4,-4);	

	\draw[fill] (-4,4) circle [radius =0.05];	
	\draw[fill] (-4,3.5) circle [radius =0.05];		
	\draw[fill] (-4,3) circle [radius =0.05];	
	\draw[fill] (-4,2.5) circle [radius =0.05];	
	\draw[fill] (-4,2) circle [radius =0.05];	
	\draw[fill] (-4,1.5) circle [radius =0.05];	
	\draw[fill] (-4,1) circle [radius =0.05];	
	\draw[fill] (-4,0.5) circle [radius =0.05];	
	\draw[fill] (-4,0) circle [radius =0.05];	
	\draw[fill] (-4,-0.5) circle [radius =0.05];	
	\draw[fill] (-4,-1) circle [radius =0.05];	
	\draw[fill] (-4,-1.5) circle [radius =0.05];	
	\draw[fill] (-4,-2) circle [radius =0.05];	
	\draw[fill] (-4,-2.5) circle [radius =0.05];	
	\draw[fill] (-4,-3) circle [radius =0.05];
	\draw[fill] (-4,-3.5) circle [radius =0.05];	
	\draw[fill] (-4,-4) circle [radius =0.05];

	\draw[fill] (12,4) circle [radius =0.05];	
	\draw[fill] (12,3.5) circle [radius =0.05];		
	\draw[fill] (12,3) circle [radius =0.05];	
	\draw[fill] (12,2.5) circle [radius =0.05];	
	\draw[fill] (12,2) circle [radius =0.05];	
	\draw[fill] (12,1.5) circle [radius =0.05];	
	\draw[fill] (12,1) circle [radius =0.05];	
	\draw[fill] (12,0.5) circle [radius =0.05];	
	\draw[fill] (12,0) circle [radius =0.05];	
	\draw[fill] (12,-0.5) circle [radius =0.05];	
	\draw[fill] (12,-1) circle [radius =0.05];	
	\draw[fill] (12,-1.5) circle [radius =0.05];	
	\draw[fill] (12,-2) circle [radius =0.05];	
	\draw[fill] (12,-2.5) circle [radius =0.05];	
	\draw[fill] (12,-3) circle [radius =0.05];
	\draw[fill] (12,-3.5) circle [radius =0.05];	
	\draw[fill] (12,-4) circle [radius =0.05];	
	
	\draw (0,4.5) node {$\mathcal{I}^{+}$};
	\draw (3.5,4.5) node {$i^{+}$};
	\draw (-4,4.5) node {$i^{+}$};
	\draw (12.5,4) node {$i^{+}$};
	\draw (-4,-4.5) node {$i^{-}$};
	\draw (0,-4.5) node {$\mathcal{I}^{-}$};
	\draw (4,-4.5) node {$i^{-}$};
	\draw (12,-4.5) node {$i^{-}$};

	\draw (4,0.3) node {\footnotesize{Static}};
	\draw (4,-0.3) node {\footnotesize{patch}};
    
    \draw (0,2.4) node {\footnotesize{patch}};
    \draw (0,3) node {\footnotesize{Milne}};
    
	\draw (8,4.3) node {\footnotesize{Black hole}};
	\draw (8,3.7) node {\footnotesize{interior I}};
    
	\draw (5.8,8.3) node {\footnotesize{Black hole}};
    \draw (5.8,7.7) node {\footnotesize{interior II}};

	\draw (1.8,2.2) node[rotate=45] {$r=r_{c}$};
	\draw (2.3,1.8) node[rotate=45] {\footnotesize{cosmological horizon}};

 	\draw (6.3,2.2) node[rotate=-45] {$r=r_{+}$};   
	\draw (5.8,1.7) node[rotate=-45] {\footnotesize{outer horizon}};

    \draw (5.8,6.3) node[rotate=45] {$r=r_-$};
	\draw (6.3,5.8) node[rotate=45] {\footnotesize{inner horizon}};
	
\end{tikzpicture}
\end{center}
\vspace{-0.7cm}
\caption{A tile of the Penrose diagram of Reissner-Nordström-de Sitter (RNdS$_4$). This tile can be continued through attaching identical tiles on their vertical dotted lines or horizontal dotted lines, in order to obtain the global causal structure. The wavy lines indicate curvature singularities and the diagonal lines indicate horizons.}
\label{fig:penrosernds}
\end{figure}
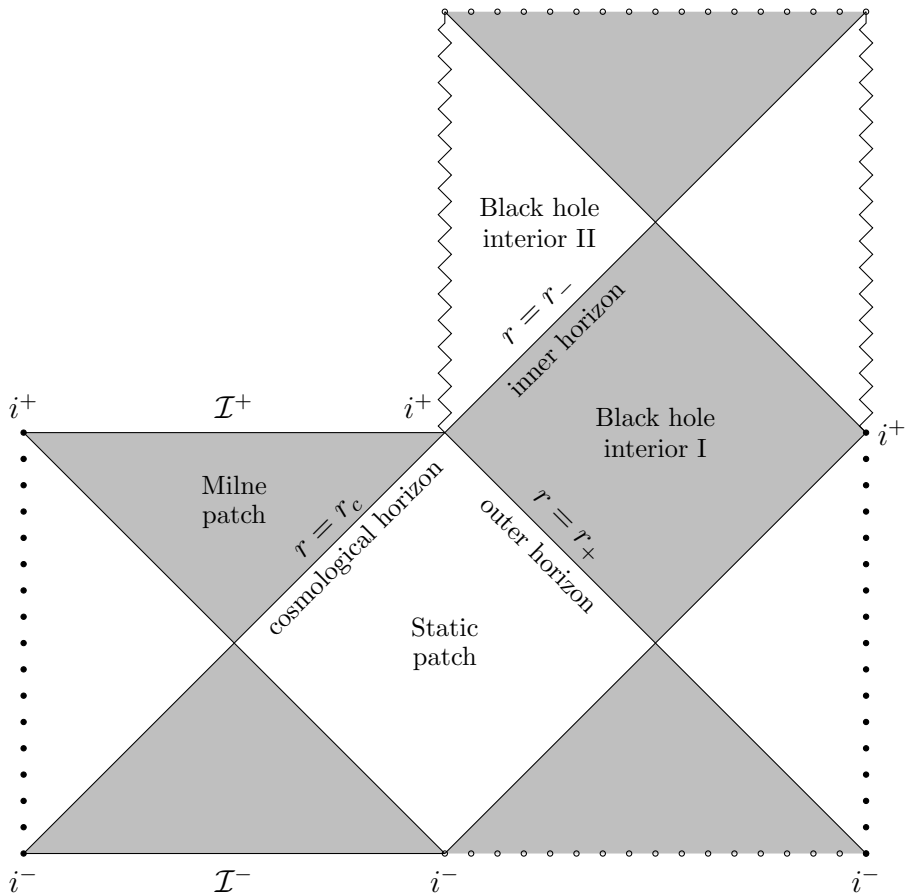

We will focus on static and spherically symmetric configurations. The first is the Reissner-Nordstr\"om solution (RNdS$_4$), which is supported by an electromagnetic field. In the Schwarzschild gauge, this solution reads
\begin{equation}\label{eq:4dvanillametric}
    ds^{2}=-V(r)\d t^{2}+\frac{\d r^{2}}{V(r)}+r^{2}d\Omega^{2}\,,
    \qquad
    d\Omega^{2}= \d\theta^{2}+\sin^{2}\theta \d\phi^{2}\,,
\end{equation}
with
\begin{equation}\label{eq:gaugefield}
    A =\frac{Q}{r}\d t\,,
    \qquad
    F=-\frac{Q}{r^{2}}\d r\wedge \d t\,\,,
\end{equation}
where for simplicity we have taken the configuration to be purely electric (the magnetic case can be treated in full analogy and leads to equivalent results). The blackening factor $V(r)$ is
\begin{align}
    V(r) = 1 - \frac{2M}{r} + \frac{Q^2}{r^2} - \frac{r^2}{\ell_4^2}
    \,,
\label{EqVr1}
\end{align}
where $M$ and $Q$ are the black hole mass and the charge parameter respectively. $V(r)$ can alternatively be written as
\begin{align}
    V(r) = - \frac{1}{\ell_4^2 r^2} \left( r + r_+ + r_- + r_c \right) \left( r - r_- \right) \left( r - r_+ \right) \left( r - r_c \right)
    \,,
\label{EqVr2}
\end{align}
where $r_c$, $r_+$, and $r_-$ are the locations of the cosmological, outer (black hole), and inner (black hole) horizons, respectively. The three distinct horizons are defined to obey $r_{c}\geq r_{+}\geq r_{-}$.  Fig.\,\ref{fig:penrosernds} depicts the Penrose diagram for RNdS$_4$, showing the four regions bounded by these horizons.

It is also useful to express the parameters $M$, $Q$, and $\ell_4$ in terms of these radii,
\begin{equation}
    \begin{aligned}
        M = ~& \frac{1}{2\ell_4^2} \left( r_+ + r_- \right) \left( \ell_4^2 - r_+^2 - r_-^2 \right)\,, \\
        Q^2 =~ & \frac{r_+ r_-}{\ell_4^2} \left( \ell_4^2 - r_+^2 - r_-^2 - r_- r_+ \right)\,, \\
        \ell_4^2 = ~& r_c^2 + r_+^2 + r_-^2 + r_- r_+ + r_- r_c + r_c r_+
        \,.
    \end{aligned}
\label{EqQuantities}
\end{equation} 
At each horizon $r_{h}=\{r_c,r_+,r_-\}$ we can evaluate the area-law, which reads
\begin{align}
    S_h = \pi r_h^2\,. 
    \label{EqAreaLaw}
\end{align}
The associated temperature and chemical potential at each horizon are 
\begin{align}
    T_h = \frac{1}{4\pi} \left| V'(r_h) \right|, \qquad \mu_h = \frac{Q}{r_h}\,.
\label{EqThmuh}
\end{align}
Finally, each horizon adheres to a first law of black hole mechanics
\begin{equation}
    \begin{aligned}
        dM = &~ - T_- dS_- + \mu_- dQ\,, \label{EqFirstLawInnerBH} \\
        dM = &~ T_+ dS_+ + \mu_+ dQ\,, \\
        dM = &~ - T_c dS_c + \mu_c dQ \,.
    \end{aligned}
\end{equation}
\begin{figure}[ht!]
	\begin{center}
		\begin{overpic}[width=0.7\textwidth]{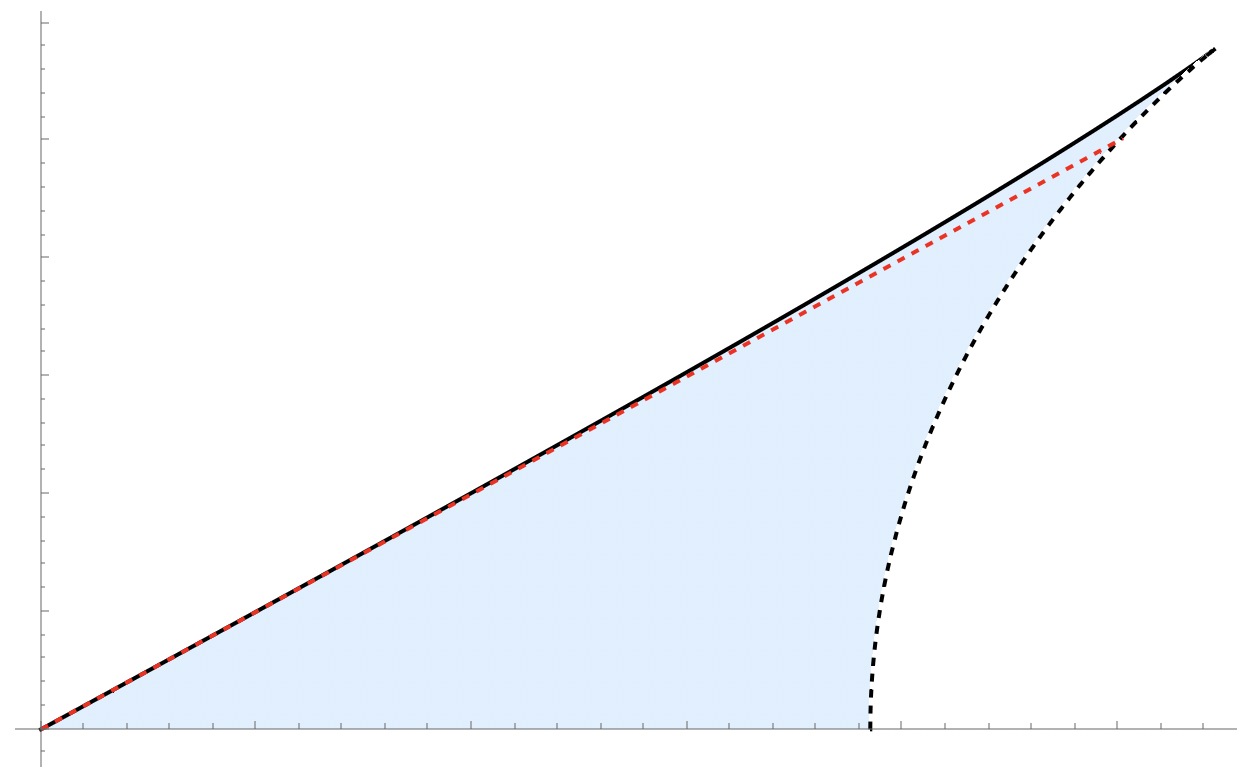}

			\put (5,58) {$|Q|$}
			
			\put (-2,12.5) {\footnotesize{0.5}}
			\put (-2,21.5) {\footnotesize{1.0}}
			\put (-2,30.5) {\footnotesize{1.5}}
			\put (-2,40.0) {\footnotesize{2.0}}
			\put (-2,49.0) {\footnotesize{2.5}}
			\put (-2,58.5) {\footnotesize{3.0}}
			
			\put (94,6) {$M$}

			\put (18,0.5) {\footnotesize{0.5}}
			\put (35.5,0.5) {\footnotesize{1.5}}
			\put (52.5,0.5) {\footnotesize{2.0}}			
			\put (69.5,0.5) {\footnotesize{2.5}}	
			\put (86.5,0.5) {\footnotesize{3.0}}

			\put (90,60) {\footnotesize{Ultracold}}
			\put (95.5,57) {$\star$}
			\put (73,45) {\rotatebox{32}{\footnotesize{Cold}}}
			\put (62,33) {\rotatebox{30}{\footnotesize{\textcolor{red}{Lukewarm}}}}
			\put (75,24) {\rotatebox{70}{\footnotesize{Nariai}}}
		\end{overpic}
	\end{center}
\vspace{-0.7cm}
\caption{The ``Shark Fin'' diagram for RNdS$_4$ which indicates allowed/disallowed combinations of $|Q|$ and $M$, for $\ell_4^2 = 100$. The shaded area corresponds to black hole solutions, and the unshaded area naked singularities. The edges correspond to the extremal black hole solutions: the solid line the Cold case, the black dashed line the Nariai case and the red dashed line highlights the lukewarm case, given by $|Q| = M$. The star indicates the Ultracold case.}
\label{fig:sharkfin}
\end{figure}
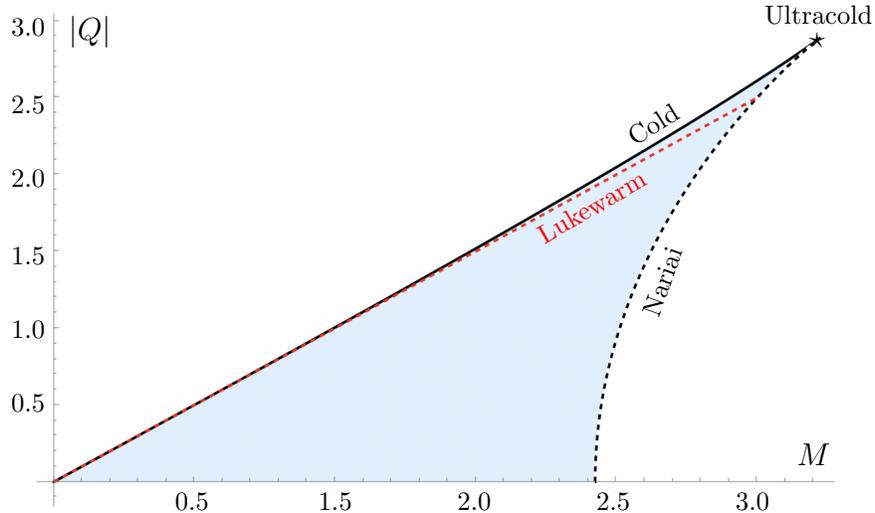
 
 The region of parameters for which the three horizons assume a real value is marked in blue in Fig.\,\ref{fig:sharkfin}, and it is usually referred to as a ``Shark Fin.'' Outside this region, we have naked singularities. The edges of the fin mark the limiting cases where two (or more) horizons coincide. These are the \textit{extremal} solutions, which are the focus of our work. The RNdS$_4$ geometry also possesses a ``lukewarm region'' where $|Q| = M$, indicated by the red dashed line in Fig.\,\ref{fig:sharkfin}. Along this line, the black hole and cosmological horizon have the same temperature $T_+ = T_c$. 
 For a theory without charged particles, any point on this line is at thermodynamic equilibrium \cite{Montero:2019ekk}. 

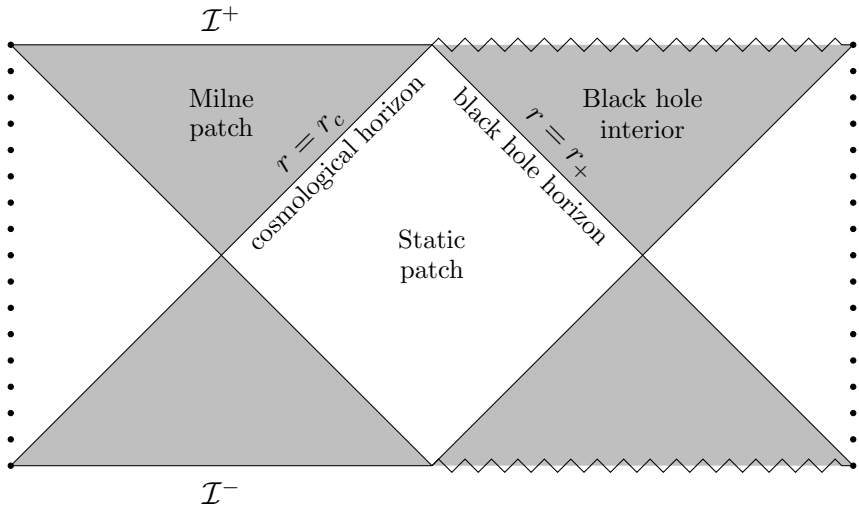
\begin{figure}[ht!]
\begin{center}
\begin{tikzpicture}[scale=0.7]

	\path [fill=lightgray] (0,0) -- (-4,4)  -- (4,4);
	\path [fill=lightgray] (0,0) -- (4,-4)  -- (-4,-4);
	
	\path [fill=lightgray] (8,0) -- (4,4)  -- (12,4);
	\path [fill=lightgray] (8,0) -- (4,-4)  -- (12,-4);

	\draw[decorate,decoration=zigzag] (4,-4)--(12,-4);
	\draw[decorate,decoration=zigzag] (4,4)--(12,4);
	
	\draw (0,0)--(4,4);
	\draw (0,0)--(4,-4);
	\draw (0,0)--(-4,-4);
	\draw (0,0)--(-4,4);
	
	\draw (4,4)--(12,-4);
	\draw (4,-4)--(12,4);

	\draw (-4,4)--(4,4);	
	
	\draw (-4,-4)--(4,-4);	

	\draw[fill] (-4,4) circle [radius =0.05];	
	\draw[fill] (-4,3.5) circle [radius =0.05];		
	\draw[fill] (-4,3) circle [radius =0.05];	
	\draw[fill] (-4,2.5) circle [radius =0.05];	
	\draw[fill] (-4,2) circle [radius =0.05];	
	\draw[fill] (-4,1.5) circle [radius =0.05];	
	\draw[fill] (-4,1) circle [radius =0.05];	
	\draw[fill] (-4,0.5) circle [radius =0.05];	
	\draw[fill] (-4,0) circle [radius =0.05];	
	\draw[fill] (-4,-0.5) circle [radius =0.05];	
	\draw[fill] (-4,-1) circle [radius =0.05];	
	\draw[fill] (-4,-1.5) circle [radius =0.05];	
	\draw[fill] (-4,-2) circle [radius =0.05];	
	\draw[fill] (-4,-2.5) circle [radius =0.05];	
	\draw[fill] (-4,-3) circle [radius =0.05];
	\draw[fill] (-4,-3.5) circle [radius =0.05];	
	\draw[fill] (-4,-4) circle [radius =0.05];	
		
	\draw[fill] (12,4) circle [radius =0.05];	
	\draw[fill] (12,3.5) circle [radius =0.05];		
	\draw[fill] (12,3) circle [radius =0.05];	
	\draw[fill] (12,2.5) circle [radius =0.05];	
	\draw[fill] (12,2) circle [radius =0.05];	
	\draw[fill] (12,1.5) circle [radius =0.05];	
	\draw[fill] (12,1) circle [radius =0.05];	
	\draw[fill] (12,0.5) circle [radius =0.05];	
	\draw[fill] (12,0) circle [radius =0.05];	
	\draw[fill] (12,-0.5) circle [radius =0.05];	
	\draw[fill] (12,-1) circle [radius =0.05];	
	\draw[fill] (12,-1.5) circle [radius =0.05];	
	\draw[fill] (12,-2) circle [radius =0.05];	
	\draw[fill] (12,-2.5) circle [radius =0.05];	
	\draw[fill] (12,-3) circle [radius =0.05];
	\draw[fill] (12,-3.5) circle [radius =0.05];	
	\draw[fill] (12,-4) circle [radius =0.05];	
	
	\draw (0,4.5) node {$\mathcal{I}^{+}$};
	\draw (0,-4.5) node {$\mathcal{I}^{-}$};

	\draw (4,0.3) node {\footnotesize{Static}};	
	\draw (4,-0.3) node {\footnotesize{patch}};	
	\draw (0,3) node {\footnotesize{Milne}};
	\draw (0,2.4) node {\footnotesize{patch}};
	\draw (8,3) node {\footnotesize{Black hole}};
	\draw (8,2.4) node {\footnotesize{interior}};

	\draw (1.7,2.15) node[rotate=45] {$r=r_c$};
	\draw (2.2,1.75) node[rotate=45] {\footnotesize{cosmological horizon}};
	\draw (6.4,2.1) node[rotate=-45] {$r=r_+$};
	\draw (5.9,1.7) node[rotate=-45] {\footnotesize{black hole horizon}};

\end{tikzpicture}
\end{center}
\vspace{-0.5cm}
\caption{A tile of the Penrose diagram of Schwarschild-de Sitter. To obtain the global causal structure, this tile can be continued through attaching identical tiles at the vertical dotted lines. The wavy lines indicate regions of strong coupling and the diagonal lines indicate horizons.}
\label{fig:penroseSdS}
\end{figure}

We will also be concerned with the Schwarzschild-de Sitter black hole solution (SdS$_4$) in Sec.\,\ref{SecSdSNariai} as the chargeless limit of the RNdS$_4$ solution. In the context of the present discussion, it corresponds to taking $Q = 0$ in \eqref{eq:4dvanillametric}-\eqref{EqVr1}, so that now we have
\begin{align}
    V(r) = 1 - \frac{2M}{r} - \frac{r^2}{\ell_4^2}\,.
    \label{EqVrSdS}
\end{align}
There are only two distinct horizons $r_+ \leq r_c$, separating three distinct regions in the spacetime, as illustrated in Fig.\,\ref{fig:penroseSdS}. 
We can rewrite the metric in terms of these horizons as
\begin{align}
    V(r) = - \frac{\left(r-r_+\right)\left(r-r_c\right)\left(r + r_++r_c\right)}{r\ell_4^2}\,,
\end{align}
where 
\begin{equation}
    \begin{aligned}
        \ell_4^2 = & r_c^2 + r_+^2 + r_c r_+\,, \\
        M = & \frac{1}{2\ell_4^2} r_+ r_c \left( r_c + r_+ \right)\,.
    \end{aligned}
\end{equation}
It will also be useful to note that 
\begin{align}
    r_c = \frac{1}{2} \left( \sqrt{4\ell_4^2 - 3 r_+^2} - r_+ \right)\,.
\label{Eqrcinrp}
\end{align}
For the SdS$_4$ solution, one can analogously define an associated temperature and first law, which are simply the $Q=0$ limits of the expressions above.

\subsection{Cold limit} \label{sec:cold}

We focus first on the \textit{Cold} black hole limit \cite{Romans:1991nq} of RNdS$_4$. In this limit the inner and outer black hole horizons coincide, that is $r_+ =r_- \equiv r_0$, while the cosmological horizon is far away.\footnote{As the cosmological horizon is far away, this configuration shares some similarities with the extremal RN black hole in flat space, which is recovered via the limit $\ell_4 \rightarrow \infty$.} The Cold limit corresponds to the diagonal line in Fig.\,\ref{fig:sharkfin}. The black hole mass $M_0$ and charge $Q_0$ in this limit are related via
\begin{align}
    Q_0^2 = r_0^2 \left( 1 - 3 \frac{r_0^2}{\ell_4^2} \right), \qquad M_0 = r_0 \left( 1 - 2 \frac{r_0^2}{\ell_4^2} \right) \,,
\label{EqColdQ0M0}
\end{align}
and the metric warp factor $V(r)$ exhibits two coincident horizons,
\begin{align}
    V_{\text{cold}}(r) = \left( 1 - \frac{r_0}{r} \right)^2 \left( 1 - \frac{r^2}{\ell_4^2} -2 \frac{r_0 r}{\ell_4^2} -3\frac{r_0^2}{\ell_4^2} \right) \,.
\end{align}
In other words, when the cosmological constant is fixed, the Cold branch is a family of solutions parameterized by the real parameter $r_0$ that satisfies
\begin{align}\label{constraint_cold}
0 \leq  r_0 < \frac{\ell_4}{ \sqrt{6}},
\end{align}
which preserves $r_0 < r_c$, i.e., the physical order of the horizons. To obtain the near-horizon geometry, we zoom in by means of the change of coordinate
\begin{align}\label{eq:decouple-cold}
r= r_0 + \lambda \rho\,, \qquad t = \frac{\ell_{\text{AdS}}^2}{\lambda} T,
\end{align}
and send $\lambda \rightarrow 0$ to obtain
\begin{align}
ds^2_{\text{NH,Cold}} = \ell_{\text{AdS}}^2 \left(-\rho^2 \d T^2 + \frac{\d \rho^2}{\rho^2} \right) +r_0^2 d\Omega^2.
\label{EqNHColdGeometry}
\end{align}
This geometry is  AdS$_2 \times S^2$, and the coordinate $\rho$ parameterises the distance from the extremal horizon. Here,
\begin{equation}\label{eq:ads2-radius}
\ell_{\text{AdS}}^2 = \frac{r_0^2{\ell_4^2}}{{\ell_4^2}-6{r_0^2}}~,
\end{equation}
is an effective AdS$_2$ radius, while the radius of the $S^2$ is $r_0$. 
We note that the constraint \eqref{constraint_cold} ensures that $\ell_{\text{AdS}}^2$ is positive.
The gauge field \eqref{eq:gaugefield} becomes
\begin{align}\label{eq:gauge-cold}
    A_0 = Q_0 \frac{\ell_{\text{AdS}}^2}{r_0} \left( \frac{1}{ \lambda} - \frac{\rho}{r_0} \right) \d T \,,
\end{align}
noting that whilst the first term is divergent in the $\lambda \rightarrow 0$, it can be removed by a gauge transformation. As all of our physics depends on $F = \d A$, it is safe to subtract this divergence, and we will not report it again.
 Finally, it will prove convenient to rewrite \eqref{EqColdQ0M0} as
\begin{align}
    Q_0^2 = \frac{r_0^2}{2\ell_{\text{AdS}}^2} \left( \ell_{\text{AdS}}^2 +r_0^2  \right), \qquad M_0 = \frac{r_0}{3\ell_{\text{AdS}}^2} \left( 2\ell_{\text{AdS}}^2+r_0^2 \right) \,.
\label{EqColdQ0M01}
\end{align}

Note that the metric \eqref{EqNHColdGeometry} is Lorentzian. Under the wick rotation $T \rightarrow - i \tau$, we obtain a (Euclidean) $\text{EAdS}_2 \times S^2$ geometry. The study of the quantum corrections on this background geometry, and the needed procedure to take this configuration out of extremality, will be the subject of Sec.\,\ref{SecColdLimit}.

\subsection{Nariai limit}\label{sec:Nariai}
The \textit{Nariai} configuration is obtained when the cosmological horizon and the outer black hole horizon coincide, that is $r_+ = r_c \equiv r_N$. The Nariai limit corresponds to the curved black dashed line in Fig.\,\ref{fig:sharkfin}. The charge and mass are constrained to be
\begin{align}
    Q_N^2 = r_N^2 \left( 1 - 3 \frac{r_N^2}{\ell_4^2} \right), \qquad M_N = r_N \left( 1 - 2 \frac{r_N^2}{\ell_4^2} \right),
\label{EqNariaiQnMn}
\end{align}
and, analogous to the Cold case, the metric warp factor $V(r)$ exhibits two coincident horizons
\begin{align}
    V_{\text{Nariai}}(r) = \left( 1 - \frac{r_N}{r} \right)^2 \left( 1 - \frac{r^2}{\ell_4^2} -2 \frac{r_N r}{\ell_4^2} -3\frac{r_N^2}{\ell_4^2} \right). 
\end{align}
The lukewarm line $|Q| = M$ intersects the Nariai configuration when $r_N = \ell_4/2$. As the Schwarzschild-de Sitter black hole also has a cosmological and black hole horizon, there is also a Nariai limit there when $Q_N = 0$. In particular, in the SdS$_4$ case we have
\begin{align}
     M_N =\frac{\ell_4}{3 \sqrt3}\,, \qquad \qquad r_N = \frac{\ell_4}{\sqrt3}\,,
\label{EqNariaiSchw}
\end{align}
and the warp factor becomes
\begin{align}
    V_{\text{Nariai},Q_N=0}(r) = - \frac{3 r^2}{r_N^2} \left( 1 - \frac{r_N}{r} \right)^2 \left( 1 + 2 \frac{r_N}{r} \right)\,. 
\end{align}
In short, we have a family of solutions parameterized by $r_N$, with the additional constraint
\begin{equation}\label{range_Nariai}
\frac{\ell_4^2}{6} < r_N^2 \leq \frac{\ell_4^2}{3}\,,
\end{equation}
where the $Q_N = 0$ saturates the upper bound of \eqref{range_Nariai}.

In what follows we will see that, by taking appropriate limits, we can obtain different patches of dS$_2 \times S^2$ near horizon geometries. The size of dS$_2$ is
\begin{equation}\label{dS2_radius}
\ell_{\text{dS}}^2 = \frac{r_N^2 \ell_4^2}{ 6r_N^2 -\ell_4^2}\,.
\end{equation}
The constraint \eqref{range_Nariai} ensures that $\ell_{\text{dS}}^2$ is positive. It will be convenient to rewrite \eqref{EqNariaiQnMn} as 
\begin{align}
    Q_N^2 = \frac{r_N^2}{2\ell_{\text{dS}}^2} \left( \ell_{\text{dS}}^2 -r_N^2  \right), \qquad M_N = \frac{r_N}{3\ell_{\text{dS}}^2} \left( 2\ell_{\text{dS}}^2-r_N^2 \right) \,.
\label{EqNariaiQnMnv2}
\end{align}
In Sec.\,\ref{SecSdSNariai} and Sec.\,\ref{SecNariaiCorrections}, we will probe the near-horizon geometry of the Nariai limit in a number of different ways. We now introduce a number of decoupling limits necessary to do so. 
\subsubsection*{\texorpdfstring{dS$_{2} \times S^2$ }{dS2xS2} geometry}
We begin by proceeding analogously to the Cold case. We zoom into the coincident horizons with the following change of coordinates
\begin{align}
\label{EqdS2decoupling}
r= r_N + \lambda \rho\,, \qquad t = \frac{\ell_{\text{dS}}^2}{\lambda} T,
\end{align}
with $\lambda \rightarrow 0$. We obtain
\begin{equation}
\begin{aligned}
ds^2_{\text{NH,Nariai}} = & \ell_{\text{dS}}^2 \left(\rho^2 \d T^2 - \frac{\d \rho^2}{\rho^2} \right) +r_N^2 d\Omega^2, \label{EqNariaiGeometry1} \\
A_{\text{NH,Nariai}} = & -Q_N \frac{\ell_{\text{dS}}^2 \rho}{r_N^2} \d T \,,
\end{aligned}
\end{equation}
a dS$_2 \times S^2$ geometry. Again, $\rho$ parameterises the distance from the extremal horizon. Observe that this corresponds to \eqref{EqNHColdGeometry} with the sign in front of the 2D metric flipped. If we choose $\lambda > 0$ and $R > 0$, the decoupling limit \eqref{EqdS2decoupling} corresponds to the Milne patch in Fig.\,\ref{fig:penrosernds}. 

To obtain an Euclidean geometry, one would instinctively make the Wick rotation $T \rightarrow i \tau$. Doing so, we obtain an overall minus sign in front of the 2D part of the metric
\begin{equation}
\begin{aligned}
ds^2 = & -\ell_{\text{dS}}^2 \left(\rho^2 \d \tau^2 + \frac{\d \rho^2}{\rho^2} \right) +r_N^2 d\Omega^2 \,,\\
A = & -i Q_N \frac{\ell_{\text{dS}}^2 \rho}{r_N^2}  \d \tau \,.
\label{EqNariaiGeometryEuclidean1}
\end{aligned}
\end{equation}
We will describe \eqref{EqNariaiGeometryEuclidean1} as a $(-\text{EAdS}_2) \times S^2$ metric, where the minus sign refers to the overall minus sign in
front of the Euclidean $\text{AdS}_2$ part of the metric. We emphasize that here we have again landed on something that looks analogous to a $\text{EAdS}_2$ geometry but with a \textit{relative} minus sign in the metric. 

One can safely obtain the Lorentzian and Euclidean near horizon geometries for SdS$_4$  by taking the $Q_N = 0$ in  \eqref{EqNariaiGeometry1} and \eqref{EqNariaiGeometryEuclidean1}. The geometry is again locally dS$_{2} \times S^2$, with a dS$_2$ radius given by $\ell_{\text{dS}}^2=r_N^2$.

\subsubsection*{\texorpdfstring{$-$(EAdS$_{2} \times S^2$) }{-(EAdS2xS2)} geometry}
In the $Q_N \not = 0$ case, one could alternatively construct a Euclidean geometry by beginning with the Lorentzian metric \eqref{EqNariaiGeometry1} and performing a \textit{complex} rotation of the radial coordinate. Explicitly, under
\begin{align}
    \rho \rightarrow i R\,, \qquad  r_N \rightarrow i R_N\,,
\end{align}
the metric becomes
\begin{equation}
\begin{aligned}
    ds^2 = & - \frac{R_N^2 \ell_4^2}{6R_N^2 + \ell_4^2} \left( R^2 \d T^2 + \frac{\d R^2}{R^2} \right) - R_N^2 d\Omega^2 \,, \label{EqMinusEAdS2S2} \\
    A = & i Q_N \frac{\ell_{\text{dS}}^2 \rho}{R_N^2} \d T \,,
\end{aligned}
\end{equation}
We write out the factor $R_N^2 \ell_4^2 /\left( 6R_N^2 + \ell_4^2 \right)$ explicitly to emphasize that it is positive for $R_N\in \mathbb{R}$. Thus, the metric \eqref{EqMinusEAdS2S2} is $- \left( \text{EAdS}_2 \times S^2 \right)$, and it is a valid solution for Einstein-Maxwell's equation with positive $\Lambda$. 

\subsubsection*{\texorpdfstring{EAdS$_{2} \times (-S^2)$ }{EAdS2x(-S2)} geometry}
It is important to note that the $- \left( \text{EAdS}_2 \times S^2 \right)$ \textit{cannot} be recovered in the SdS$_4$ case. In particular, in the $Q_N= 0$ limit, $\ell_4^2 = 3 r_N^2 = - 3 R_N^2$. In this case, the metric becomes
\begin{align}
    ds^2 =  R_N^2 \left( R^2 \d T^2 + \frac{\d R^2}{R^2} \right) - R_N^2 d\Omega^2.
\label{EqMinusEAdS2S2Q0}
\end{align}
Specifically, we have landed on a $\text{EAdS}_2 \times \left( - S^2 \right)$. That is, we have a relative minus sign rather than an overall minus sign. The metric \eqref{EqMinusEAdS2S2Q0} therefore looks more analogous to \eqref{EqNariaiGeometryEuclidean1}, but the complexification of the radial coordinate has just swapped its location.

\subsubsection*{\texorpdfstring{$S^{2} \times S^2$ }{S2xS2} geometry}
Finally, it is worth noting that one could have also obtained $\text{dS}_2 \times S_2$ in static patch coordinates. Beginning at \eqref{EqNariaiGeometry1} and making the coordinate redefinitions
\begin{align}
    \rho = i \cos \hat{\theta} - \sin \hat{\theta} \cosh \hat{t}, \qquad \text{ and } \qquad T = - i \frac{\sin \hat{\theta} \sinh \hat{t}}{i \cos \hat{\theta} - \sin \hat{\theta} \cosh \hat{t}},
\end{align}
the metric becomes
\begin{equation}
\begin{aligned}
    ds^2 = & \ell_{\text{dS}}^2 \left( \d \hat{\theta}^2 - \sin^2 \hat{\theta} \d \hat{t}^2 \right) + r_N^2 d\Omega^2 \,, \label{eqstaticpatchcoord} \\
    A = &\, iQ_N \frac{\ell_{\text{dS}}^2}{r_N^2}  \cos\hat\theta\, \d\hat{t}\, .
\end{aligned}
\end{equation}
Here we have written the resulting gauge field after removing a pure gauge contribution.

The Lorentzian metric \eqref{eqstaticpatchcoord} describes the static patch \cite{Anninos:2012qw}. We obtain a Euclidean metric via the Wick rotation $\hat{t} \rightarrow i \hat{\phi}$, specifically 
\begin{align}
    ds^2 = & \ell_{\text{dS}}^2 \left( \d \hat{\theta}^2 + \sin^2 \hat{\theta} \d \hat{\phi}^2 \right) + r_N^2 d\Omega^2\,, \\
    A = &\, -Q_N \frac{\ell_{\text{dS}}^2}{r_N^2}  \cos\hat\theta\, \d\hat{\phi}\, .
\end{align}
We recognise this Euclidean geometry as an $S^2 \times S^2$ with radii $\ell_{\text{dS}}$ and $r_N$, respectively.

\subsection{Ultracold limit}
\label{SecUCGeometry}
So far, we have only considered instances where two horizons are coincident. For completeness, (when $Q \not = 0$) we note that there is also an instance where all three horizons coincide, $r_+ = r_- = r_c \equiv r_{\text{uc}}$. This limit is known as the \textit{Ultracold} solution and corresponds to the tip of the Shark Fin, marked by a star in Fig.\,\ref{fig:sharkfin}. In this case, the mass and charge are 
\begin{align}
    Q_{\text{uc}}^2 = \frac{\ell_4^2}{12}\, , \qquad M_{\text{uc}} = \frac{\sqrt2 \ell_4}{3 \sqrt3}\,,  \qquad r_{\text{uc}} = \frac{\ell_4}{\sqrt{6}}\, . 
\label{EqNariaiQucMuc}
\end{align}
The metric warp factor exhibits three coincident horizons
\begin{align}
    V_{\text{uc}}(r) = & - \frac{r^2}{6r_{\text{uc}}^2} \left( 1 - \frac{r_{\text{uc}}}{r} \right)^3 \left( 1 + 3 \frac{r_{\text{uc}}}{r} \right)\,.
\end{align}
By slightly displacing the horizons via $r_c = r_{\text{uc}} + \lambda$ and $r_0 = r_{\text{uc}} - \lambda$, and subsequently taking the decoupling limit
\begin{align}
   r = r_{\text{uc}} +  \sqrt{\frac{2}{3 r_{\text{uc}}^3}}\lambda^{3/2} x\, ,  \qquad t = \sqrt{\frac{3 r_{\text{uc}}^3}{2}} \frac{\tau}{\lambda^{3/2}}\,,
\label{EqNariaidecouple}
\end{align}
we arrive at the $\text{Mink}_2 \times S^2$ near horizon geometry,
\begin{equation}
\begin{aligned}
  ds^2 = & -\d \tau^2 + \d x^2 + r_{\text{uc}}^2 d\Omega^2\,, \label{EqUCnMn} \\
    A = & -\frac{Q_{\text{uc}} x}{r_{\text{uc}}^2} d\tau \, .
\end{aligned}
\end{equation}
The treatment of the Ultracold case turns out to be quite challenging, for reasons which we will elaborate upon in Sec.\,\ref{sec:conclusion}. In this paper, we will focus on studying in-depth quantum effects in the Cold and Nariai limits.

\section{Aspects of Euclidean path integrals near to extremality}
\label{Sec3GenStrat}
In this section we review how to obtain the dominant correction to the saddle-point approximation of the Einstein-Maxwell action (with a positive cosmological constant) near extremality, using a Euclidean path integral approach. 
We first explicitly identify the dominant corrections in Sec.\,\ref{subsec:correct}, before discussing our strategy for computing them in Sec.\,\ref{subsubtensor} and Sec.\,\ref{sec:away}. The discussion of this section follows from \cite{Iliesiu:2022onk,Banerjee:2023quv}.

\subsection{Saddle-point approximation and quantum corrections}
\label{subsec:correct}
The Euclidean path integral is given formally by an integral over all possible geometries and fields weighted by the appropriate action, i.e.,  
\begin{align}
    Z = \int \left[ Dg \right]\left[DA\right] e^{-I[g,A]}\,,\qquad  I[g,A] 
    = 
    I_{\text{EM}} 
    + 
    I_{\text{boundary}}
    +
    I_{\text{gauge}}
    \,.
\end{align}
Here, $I_{\text{EM}}$ represents the wick-rotated Einstein-Maxwell action introduced in \eqref{eq:vanilla4daction}, depending on metric $g_{\mu\nu}$ and gauge field $A_{\mu}$. The term $I_{\text{boundary}}$ imposes Dirichlet and Neumann boundary conditions on the metric and gauge field, respectively. 
Finally, the term $I_{\text{gauge}}$ will ensure the gauge fixing of the metric and gauge field.

We will evaluate the path integral via a saddle-point approximation, where the saddles we will focus on are the extremal and near-extremal solutions in Sec.\,\ref{sec:cold} and Sec.\,\ref{sec:Nariai}. Around each saddle 
\begin{align}
    Z \approx Z_0 = \exp \left( - I[\bar{g},\bar{A}] \right)\,.
\end{align}
Here $\bar{g}$ and $\bar{A}$ denote a classical solution to the equations of motion of the Einstein-Maxwell action.
The goal of this work is to compute the leading quantum contributions to the path integral as one heats up the extremal solutions and contrast these quantum effects to their classical counterpart.

Computationally, we expand the action $I[g,A]$ about the saddle-point $(\bar g, \bar A)$, where  
\begin{equation}
g = \bar{g} + h\,,\qquad A = \bar{A} + \frac{1}{2} a\,, 
\end{equation}
and $h$ and $a$ are quantum fluctuations. The first order term in $a$ and $h$ vanishes as the background  $(\bar g, \bar A)$ satisfies the classical equations of motion.   Our choice of gauge for these fluctuations reads
\begin{equation}
    \begin{aligned}
    \mathcal{L}_{\text{gauge, diffeo}} =~ & \frac{1}{32 \pi} \bar{g}_{\mu \nu} \left( \bar{\nabla}_\alpha h^{\alpha \mu} - \frac12 \bar{\nabla}^\mu h^{\alpha \;} _{\;\alpha} \right) \left( \bar{\nabla}_\beta h^{\beta \nu} - \frac12 \bar{\nabla}^\nu h^{\beta \;} _{\; \beta} \right) \label{EqGuageTT}, \\
    \mathcal{L}_{\text{gauge, }U(1)} =~ & \frac{1}{32 \pi} \left( \bar{\nabla}_{\alpha} a^{\alpha} \right)^2
    \,,
\end{aligned}
\end{equation}
where the bars denote covariant derivatives with respect to the classical background geometry.
The metric gauge fixing term $\mathcal{L}_{\text{gauge, diffeo}}$ specifically enforces the transverse-traceless gauge. The photon gauge fixing term $\mathcal{L}_{\text{gauge, }U(1)}$ corresponds to Lorenz gauge.
The leading quantum contribution comes from quadratic terms in the expansion of $I[g,A]$.
This is the one-loop term in the effective action, and it is controlled by the linearized kinetic operators $D$ and $P$, also referred to as generalized Lichnerowicz operators, as well as a mixing term $O_{\text{int}}$. In short, the path integral up to this quadratic order  takes the form
\begin{align}
    \begin{split}
        Z \approx \exp \left( - I [\bar{g},\bar{A}] \right) \int [Dh][Da] \exp & \Big[ - \int d^4 x \sqrt{\bar{g}} \Big(  h^* D[\bar{g},\bar{A}] h   \\
    &  \quad + a^* P[\bar{g},\bar{A}]a + \left( h^* O_{\text{int}}[\bar{g},\bar{A}]a + \text{h.c} \right) \Big) \Big]
    \,,
    \end{split}
\label{EqZGaussian1}
\end{align}
where the star denotes a complex conjugate,\footnote{Note that one could choose to either a) construct complex modes, or b) separate out into real and imaginary modes. We choose the former; both are equivalent, provided the action itself is real.} and this expression includes the gauge fixing terms in \eqref{EqGuageTT}. For our situation, Einstein-Maxwell theory with a  cosmological constant, these operators are \cite{Christensen:1979iy,Maulik:2024dwq}
\begin{equation}\begin{aligned}
         &h_{\alpha \beta}^* \,D^{\alpha \beta, \mu \nu} [\bar{g}]\, h_{\mu \nu} =\\
         & -\frac{1}{16 \pi} h_{\alpha \beta}^* \bigg( \frac14 \bar{g}^{\alpha \mu} \bar{g}^{\beta \nu} \bar{\Box} -\frac18 \bar{g}^{\alpha \beta} \bar{g}^{\mu \nu} \bar{\Box} +\frac12 \bar{R}^{\alpha \mu \beta \nu} +\frac12 \bar{R}^{\alpha \mu} \bar{g}^{\beta \nu} \\
         &  - \frac12 \bar{R}^{\alpha \beta} \bar{g}^{\mu \nu} -\frac14 \bar{R} \bar{g}^{\alpha \mu} \bar{g}^{\beta \nu} + \frac18 \bar{R} \bar{g}^{\alpha \beta} \bar{g}^{\mu \nu} +\frac18 \bar F^2 \big(2 \bar g^{\alpha \mu} \bar g^{\beta \nu} - \bar g^{\alpha \beta} \bar g^{\mu \nu} \big) \\
         & -\bar F^{\alpha \mu} \bar F^{\beta \nu} - 2 \bar F^{\alpha \gamma} {\bar F^{\mu}}_{\gamma} \bar g^{\beta \nu} + \bar F^{\alpha \gamma} {\bar F^{\beta}}_{\gamma} \bar g^{\mu \nu} -\frac{\Lambda}{2} \bar g^{\alpha \mu} \bar g^{\beta \nu} + \frac{\Lambda}{4} \bar g^{\alpha \beta} \bar g^{\mu \nu}    \bigg) h_{\mu \nu} \;,
\label{EqLinGravOperator}
\end{aligned}\end{equation}
and
\begin{align}
    a_{\mu}^* \,P^{\mu \nu}\, a_{\nu} = -\frac{1}{32\pi}a_{\mu}^* \left( \bar{g}^{\mu \nu} \bar\Box - \bar{R}^{\mu \nu}  \right) a_{\nu}\,.
\label{EqPhotonOperator}
\end{align}
The mixing term reads
\begin{align}
    h_{\alpha \beta}^*\, O_{\text{int}}^{\alpha \beta \mu}\, a_{\mu} = \frac{1}{16\pi} h_{\alpha \beta}^* \left( 4 \bar{g}^{\alpha [\mu} \bar{F}^{\nu] \beta} +\bar{F}^{\mu \nu} \bar{g}^{\alpha \beta} \right) \nabla_{\mu} a_{\nu}\,.
\label{EqIntha}
\end{align}
 We note that in \eqref{EqZGaussian1} there is also a contribution from a ghost Lagrangian for ghost fields $\lbrace b_{\mu}, c_{\mu} \rbrace$ associated with diffeomorphism invariance and $\lbrace b^{(i)}, c^{(i)} \rbrace$ associated with $U(1)$ gauge invariance. Here the index $(i)$ accounts for the possibility of multiple gauge fields, although we are only concerned with one in this work. The contribution of all these ghosts is \cite{Christensen:1979iy,Banerjee:2010qc} 
\begin{align}
    \mathcal{L}_{\text{Ghost}} = - \frac{1}{16\pi} \left( b_{\mu} \left( \bar{g}^{\mu \nu} \bar{\Box} + \bar{R}^{\mu \nu} \right) c_{\nu} + b^{(i)} \bar{\Box}c^{(i)} - 2 b^{(i)} \bar{F}^{\mu \nu} \bar{\nabla}_{\mu} c_{\nu} \right)\,.
\label{EqLagGhost}
\end{align}

The outcome of performing the path integral \eqref{EqZGaussian1} is a standard Gaussian integral, with the resulting expression being controlled by the determinants of the operators appearing in the Lagrangian. In this context, the task is to have a characterisation of the eigenvalues and eigenmodes associated with $D$, $P$, $O_{\text{int}}$ and ghosts. Some of the complications that might appear in this computation are two-fold: infinitely many eigenvalues in these determinants are zero or negative. Both of these problems will appear in what follows, but we will focus first on the issues related to having eigenmodes with zero eigenvalues, i.e., zero modes. 

The appearance of a zero eigenvalue in the operators entering in \eqref{EqZGaussian1} is usually associated with a background symmetry that is not removed by gauge fixing. For example, if the background geometry $\bar g$ has a Killing vector, this will not be removed by the transverse-traceless gauge, leading to a mode with a zero eigenvalue. In this context, it is traditional to write \eqref{EqZGaussian1} as
\begin{equation}\label{eq:Zgrav1}
    Z \approx \exp \left( - I [\bar{g},\bar{A}] \right) Z_{\text{z.m.}}\, Z_{\text{n.z.m.}}~,
\end{equation}
where 
\begin{equation} \label{eq:zm}
    Z_{\text{z.m.}} = \prod_{\text{zero-modes}} \int [Dh]_{\text{zero-modes}} \int [Da]_{\text{zero-modes}}
\end{equation}
is the contribution from modes with zero eigenvalues, and $Z_{\text{n.z.m.}}$ are the contribution from eigenmodes of $D$, $P$, $O_{\text{int}}$ and ghosts with  non-zero eigenvalues. In many situations, the number of zero-modes is finite, and hence $Z_{\text{z.m.}}$ is merely a normalisation that needs to be incorporated in the answer. See, for example, \cite{Sen:2012cj,Sen:2012dw} which gives an account for this normalisation.

However, zero modes play a dominant role in the near-horizon geometry of extremal black holes. A key observation in \cite{Iliesiu:2022onk}, which builds on the developments in \cite{Charles:2019tiu,Iliesiu:2020qvm,Heydeman:2020hhw}, is that geometries of the form AdS$_2\times S^n$ will have an infinite set of zero modes due to diffeomorphism that acts on the AdS$_2$ geometry, as well as diffeomorphisms that deform the sphere, which are not removed by gauge-fixing. This implies that \eqref{eq:zm} is ill-defined, $Z_{\text{z.m.}}\to \infty$, and therefore we do not have a handle on \eqref{eq:Zgrav1} as it stands.      

A practical way to regulate this divergence is to deform the background solution slightly away from extremality by heating up the geometry \cite{Iliesiu:2022onk}.  Turning on slightly the temperature of the black hole corrects the eigenvalues of the zero modes, and rather remarkably this quantum correction is the leading effect in the path integral at low temperatures. This is the effect that we want to quantify for the extremal solutions in Sec.\,\ref{sec:dSRN}, and in the remainder of this section we will describe how the regulation procedure works in our examples. This follows closely the analysis of \cite{Banerjee:2023quv}, with appropriate modifications to incorporate a four-dimensional cosmological constant.

\subsection{Characterization of zero modes}\label{subsubtensor}

In this portion, we will describe the general properties of the zero modes that will appear in our analysis of the path integral for RNdS$_4$ black holes near extremality. Each of our extremal limits in  Sec.\,\ref{sec:dSRN} has a geometry of the form $\mathcal{M}\times S^2$, where $\mathcal{M}$ is a two-dimensional Euclidean non-compact geometry, and supported by a field strength on $\mathcal{M}$. It will be convenient to write these backgrounds as
\begin{equation}
    \begin{aligned}\label{eq:background}
        \bar g_{\mu\nu} \d x^\mu \d x^\nu &= \bar g_{ab} \d x^b \d x^a +\bar g_{ij} \d x^i \d x^j ~,\\
        \bar A&=\bar A_a \d x^a~, 
    \end{aligned}
\end{equation}
where indices $a,b$ run over the two-dimensional space $\mathcal{M}$, with background metric $\bar g_{ab}$, and the indices $i,j$ run over $S^2$, with background metric $\bar g_{ij}$. 

The zero modes appear due to diffeomorphisms that act on the boundary of $\mathcal{M}$, which annihilate $D$ in \eqref{EqLinGravOperator}, or due to $U(1)$ gauge transformations, which annihilate $P$ in \eqref{EqPhotonOperator}. This will lead to three classes of zero modes: (i) tensor modes, which are diffeomorphisms on $\mathcal{M}$, (ii) vector modes, which correspond to diffeomorphism that deform $S^2$ along $\mathcal{M}$, and (iii) gauge (photon) modes, which correspond to $U(1)$ gauge transformations acting on $\bar A$. The set of diffeomorphisms and gauge transformations that can be categorized as zero modes, and hence enter in the path integral, need to satisfy two key conditions. First, they have to comply with our choice of gauge in \eqref{EqGuageTT}. Second, the resulting fields generated by these transformations have to be smooth and single-valued, leading to a finite inner product.\footnote{For diffeomorphism acting only on $S^2$, which is a compact space, these conditions are only solved by the three Killing vectors of the sphere. Hence we will not encounter an infinite tower of zero modes for that portion of the geometry.} In the following, we will describe these criteria more explicitly for \eqref{eq:background}.  

It is important to note that the mixing term \eqref{EqLagGhost} does not give additional zero modes, which can be checked for the explicit background in play. Also, the ghost Lagrangian \eqref{EqLagGhost} has positive definite eigenvalues, and hence does not contribute to our analysis of zero modes; this is more delicate to see, which we discuss in App.\,\ref{app:ghost}.

\paragraph{(i) Tensor modes.}
Let us first consider a metric fluctuation $h$ generated by some diffeomorphism $\zeta$ on $\mathcal{M}$, where 
\begin{align}
    h_{ab} = \mathcal{L}_{\zeta} \bar{g}_{ab}~, \qquad \zeta=\zeta^a\partial_a~. 
\label{EqDiffeo}
\end{align}
 Our first restriction on $\zeta$ is that the resulting metric fluctuation $h_{ab}$ has to satisfy our transverse-traceless gauge condition \eqref{EqLinGravOperator}, which reads 
\begin{align}
    \bar{g}^{ab} h_{ab} = 0~,\quad \text{ and } \quad \bar{\nabla}^{a} h_{ab} = 0~.
\end{align}
The traceless condition is immediately satisfied if we define our vector $\zeta$ in terms of a scalar field $\Phi(x^a)$
\begin{align}
    \zeta^{a} = \epsilon^{ab} \nabla_{b} \Phi~,
\label{EqZetainPhi}
\end{align}
with $\epsilon_{ab}$ the Levi-Civita tensor on $\mathcal{M}$. The transverse condition is satisfied if the field $\Phi(x^a)$ obeys
\begin{align}
    \left( \Box_{\mathcal{M}} + R^{(2)} \right) \Phi = 0~,
\label{EqTransversecond}
\end{align}
where $R^{(2)}$ is the Ricci scalar of $\mathcal{M}$. With this, for a given $\mathcal{M}$, we can construct metric fluctuations by constructing a basis of solutions to \eqref{EqTransversecond}, which we label as $\Phi_{(k)}$, each leading to a metric perturbation $h_{ab}^{(k)}$. 

Meanwhile we require the perturbations $h^{(k)}_{ab}$ themselves to be normalisable and regular. Concretely, we require the inner product $\langle h^{(k)} | h^{(k')}\rangle$ to not be infinite nor zero, where
\begin{align}\label{eq:defn-norm}
   \langle h^{(k)} | h^{(k')}\rangle \equiv \int \d^4 x \sqrt{\bar{g}} \, ({h_{\mu \nu}}^{(k)})^* {\bar{g}}^{\mu \alpha}{\bar{g}}^{\nu \beta} h_{\alpha \beta}^{(k')} ~.
\end{align}
It is worth highlighting that this condition does not require the norm of $\zeta^a$ in \eqref{EqZetainPhi}, nor of $\Phi_n$, to be finite. It will be the case that at the boundary of $\mathcal{M}$ we will have $\zeta^a \zeta_a \to \infty$.\footnote{In \cite{Banerjee:2023quv,Iliesiu:2022onk} these are referred to as ``large diffeomorphisms.''} 

 The class of metric fluctuations $h_{ab}^{(k)}$ constructed this way are called \textit{tensor modes}. They have the property of being zero modes of the operator $D$ in \eqref{EqLinGravOperator}. As we will see in our examples, the fact that $\mathcal{M}$ is not compact, will lead to an infinite tower of these modes.

\paragraph{(ii) Vector modes.} Another set of diffeomorphisms one could consider are those generated by the vector
\begin{equation}\label{eq:diff-vec-modes}
    \xi = \Phi^0(x^a)\, \hat\xi^i\, \partial_i~,  
\end{equation}
where $\Phi^0$ is a function with support on $\mathcal{M}$ only, and $\hat\xi^i$ is a Killing vector of $S^2$ which we will write as
\begin{equation}\label{eq:killingS2}
    \hat\xi^i= \epsilon^{ij}\partial_j Y_{1,m}~,\quad m=-1,0,1~,
\end{equation}
with $Y_{l,m}(\theta,\phi)$ a spherical harmonic function and $\epsilon_{ij}$ is the Levi-Civita tensor over the $S^2$ normalised as $\epsilon_{\theta \phi} = \sin \theta$. The resulting metric fluctuation, created by acting with $\xi$ on the background, only has non-trivial components
\begin{equation}
\begin{aligned}\label{eq:vector-mode}
    h_{ai} &= \mathcal{L}_\xi \bar g_{ai} \\
    &= \bar g_{ij} \hat\xi^j \, \partial_{a} \Phi^0~,
\end{aligned}
\end{equation}
which by design is traceless. The nomenclature \textit{vector} simply comes since the metric fluctuation is a composition of vectors on $S^2$ and $\mathcal{M}$. We also need to enforce that $\nabla^\mu h_{\mu\nu}=0$, which translates in this case to the requirement 
\begin{equation}\label{EqLargeGauge}
    \Box_{\mathcal{M}}\Phi^0=0~.
\end{equation}
With this, we can construct fluctuations by solving \eqref{EqLargeGauge}, and label each independent solution by $\Phi^0_{(n)}$. Then these modes will be parametrized as $v_{ai}^{(m,n)}$, where $m$ corresponds to the choice of Killing vector in \eqref{eq:killingS2}, and $n$ a choice of $\Phi^0_{(n)}$. Finally, we also require the norm of $v_{ai}^{(m,n)}$, as defined in \eqref{eq:defn-norm}, to be neither infinite nor zero for it to be counted as a zero mode in our path integral.  With these requirements in place,  the vector modes are zero modes of $D$ in \eqref{EqLinGravOperator}.

\paragraph{(iii) Gauge (photon) modes}
\label{subsubGauge}
Finally, we consider zero modes that appear in the photon operator \eqref{EqPhotonOperator}. These are $U(1)$ gauge transformations acting on $\bar A$, and their form is very similar to the vector modes encountered above. 
In particular, we find that the only relevant photon modes to be 
\begin{align}
    a^{(k)}_{\mu} = \partial_{\mu} \Phi^0_{(k)}~,
\label{EqAmugeneral}
\end{align}
up to a constant, which is determined by the particular limit of interest. Here, $\Phi^0_{(k)}$ are the scalars satisfying \eqref{EqLargeGauge} which enforces the Lorenz gauge condition in \eqref{EqGuageTT}.  We will require that these modes have a finite (and non-zero) norm, which is defined as 
\begin{align}
\langle a^{(k)} | a^{(k')}\rangle \equiv \int \d^4 x \sqrt{\bar{g}} \, {a^{(k)}_{\mu}}^* {\bar{g}}^{\mu \nu} {a^{(k')}_{\nu}} ~.
\end{align}
For these modes, the mixing term \eqref{EqIntha} vanishes, which means that we do not have a mixing between zero modes due to diffeomorphisms and gauge transformations.

\subsection{Departing away from extremality}\label{sec:away}
In the previous subsection, we have characterized the problematic zero modes of the path integral at extremality where the background geometry is $\mathcal{M}\times S^2$. Here we want to quantify what happens to these modes by moving away from extremality. A natural way to deform away from extremality is to heat up the geometry, but the strategy applies broadly. Concretely, consider deforming \eqref{eq:background}
\begin{equation}
    \bar g \to \bar g + \delta g ~,\qquad \bar A \to \bar A +\delta A~,
\end{equation}
where $\delta g$ and $\delta A$ are linearized solutions to the equations of motion.\footnote{If this corresponds to turning on the temperature $T$, then $\delta g\sim O(T)$ and $\delta A\sim O(T)$.} This deformation will induce a modification to the operators appearing in \eqref{EqZGaussian1}, 
\begin{equation}
\begin{aligned}
     D[\bar g , \bar A] &\to D[\bar g +\delta g, \bar A + \delta A]= \bar D  + \delta D+\cdots~, \\
     P[\bar g , \bar A] &\to P[\bar g +\delta g, \bar A + \delta A]= \bar P  + \delta P+\cdots~,
\end{aligned}
    \end{equation}
where $\bar D \equiv  D[\bar g , \bar A] $ and  $\bar P \equiv  P[\bar g , \bar A] $. In the following, we will describe how the eigenstates and eigenvalues are modified due to this deformation. First, let us focus on the operator $D$: we define the eigenstates $h^{(n)}$ and eigenvalues $\bar \Lambda_n$ at extremality as
\begin{equation}
    \bar D  h^{(n)} = \bar \Lambda_n h^{(n)}~,
\end{equation}
where indices are omitted for simplicity. It is assumed that $h^{(n)}$ are orthogonal with respect to the inner product \eqref{eq:defn-norm}, i.e., $\langle h^{(n)} | h^{(m)}\rangle = \kappa_n \delta_{nm}$ with $\kappa_n$ a constant. When heating up, we will have
\begin{align}
    \left( \bar{D} + \delta D \right) \left( h^{(n)} + \delta h^{(n)}  \right) = \left( \bar \Lambda_n + \delta \Lambda_n \right) \left( h^{(n)}  + \delta h^{(n)}  \right)~,
\end{align}
which includes corrections to the eigenvalues $\delta \Lambda_n$ and to the eigenstates $\delta h^{(n)}$. To leading order in the deformation, we find
\begin{align}
\bar{D} \delta h^{(n)} + \delta D h^{(n)} = \bar\Lambda_n \delta h^{(n)} + \delta \Lambda_n h^{(n)}\,.
\end{align}
Applying an inner product of this equation against $(h^{(m)})^*$, and using orthogonality, gives
\begin{align}\label{eq:correc-D}
    \delta \Lambda_n  = \frac{1}{\kappa_n}\langle h^{(n)} | \delta D h^{(n)}\rangle = \frac{1}{\kappa_n} \int \d^4 x \sqrt{\bar{g}} \, {h^{(n)}}_{\alpha \beta}^* \delta D^{\alpha \beta \mu \nu} {h^{(n)}}_{\mu \nu}~,
\end{align}
where we also used that $\bar D$ is self-adjoint, $\langle h^{(n)} |  \bar D h^{(m)}\rangle= \langle h^{(m)} |  \bar D h^{(n)}\rangle ^*$. Following an analogous derivation, we find that the correction to the eigenvalues of $P$ reads
\begin{align}\label{eq:correc-P}
    \delta \Lambda_n = \frac{1}{\gamma_n}\int \d^4 x \sqrt{\bar{g}}\, {a^{(n)}}_{\mu}^* \delta P^{\mu \nu} {a^{(n)}}_{ \nu}~,
\end{align}
where $\gamma_n$ is the normalisation of the $U(1)$ modes, $\langle a^{(n)} | a^{(m)}\rangle = \gamma_n \delta_{nm}$. It is important to note here that the operator $O_{\text{int}}$ does generically contribute to the corrections and we should have diagonalized  $D$, $P$ and $O_{\text{int}}$ jointly (which is straightforward, but tedious). However, we will only focus on zero modes and our background \eqref{eq:background} is a direct product; for this reason $O_{\text{int}}$ does not play a role in our analysis, hence this simpler presentation applies. In more complicated situations this will no longer be true and we refer to \cite{Kolanowski:2024zrq}.

The correction to the eigenvalues has its most striking effect on zero modes.  At extremality, their path integral is given by \eqref{eq:zm}, where $\bar \Lambda_n=0$ by definition. Near-extremality, the first correction $\delta \Lambda_n$ restores the Gaussian integral and gives\footnote{Note that we are assuming here that $\kappa_n$ and $\gamma_n$, the normalisation of the modes, are positive and the corrected eigenvalue as well. This will not always be the case, and we will discuss that issue carefully as it arises. }
\begin{align}\label{eq:Zcorr}
    \delta \log Z_{\text{z.m.}}  = - \frac{1}{2} \sum_n \log \left( \delta \Lambda_n \right)~. 
\end{align}
This shows that deforming the background in principle regulates the divergent nature of $ Z_{\text{z.m.}}$ provided the eigenvalues are corrected ($\delta\Lambda_n\neq0$). As we will see in the explicit cases considered in the next sections, \eqref{eq:Zcorr} is the leading effect in the path integral as one moves away from extremality. We note that there is also a finite contribution that corrects also $Z_{\text{n.z.m.}}$, but it is smooth as we go back to extremality and therefore we will not consider this portion further.

\section{Path integral for the near-Cold black hole}
\label{SecColdLimit}

In this section, we start by analysing the path integral around the Cold limit of RNdS$_4$ described in Sec.\,\ref{sec:cold}. We will be particularly interested in the role of zero modes in this path integral, and therefore we will see how the general structure discussed in Sec.\,\ref{Sec3GenStrat} applies in this case. The Cold black hole corresponds to the limit of $r_+=r_-\equiv r_0$ and its near-horizon geometry is AdS$_2\times S^2$, as described in \eqref{EqNHColdGeometry}. For this reason, many aspects of the analysis here will follow from \cite{Iliesiu:2022onk,Banerjee:2023quv}, and we will highlight the differences due to the presence of a positive cosmological constant.  
With this in mind, we will start this section with a description of the near-extremal solution around the Cold background in  Sec.\,\ref{Seccolddecoupling}, and then discuss the role of zero modes in the extremal and near-extremal path integral in Sec.\,\ref{sec:zero-cold}. 

This extremal and near-extremal configuration has a very natural counterpart in AdS$_4$. Hence we briefly report on the appropriate results when the RN black hole is embedded in a theory with a negative cosmological constant in Sec.\,\ref{AdS4_subs}.

\subsection{Near-extremal background}
\label{Seccolddecoupling}

In this portion we will quantify the modification of the decoupling limit in Sec.\,\ref{sec:cold} to incorporate a deviation away from extremality. We deviate from extremality by slightly turning on $T_+$, as defined in \eqref{EqThmuh}, which produces an increase in $r_+$ and decrease in $r_-$ from their extremal value $r_0$.  As the temperature is turned on, we will choose to work in a canonical ensemble, where the charge $Q$ is fixed to its extremal value in \eqref{EqColdQ0M0}; we will also require that the four-dimensional de Sitter radius $\ell_4$ is fixed. With this, the change of the inner and outer horizons reads 
\begin{equation}
    \begin{aligned}
        r_- = ~& r_0 - 2 \pi \ell_{\text{AdS}}^2 T_+ - \frac{2\pi^2 \ell_{\text{AdS}}^4 }{3r_0^3}\left( 2\ell_{\text{AdS}}^2 + 7r_0^2 \right) T_+^2 + \mathcal{O} \left( T_+^3 \right)\,, \\
        r_+ = ~& r_0 + 2 \pi \ell_{\text{AdS}}^2 T_+ + \frac{2\pi^2 \ell_{\text{AdS}}^4 }{3r_0^3}\left( 2\ell_{\text{AdS}}^2 + 3r_0^2 \right) T_+^2 + \mathcal{O} \left( T_+^3 \right)\,,
    \end{aligned}
\label{Eqcolddecoupling}
\end{equation}
where the AdS$_2$ radius  $\ell_{\text{AdS}}$ is defined in \eqref{eq:ads2-radius}. 
To check that we have correctly displaced the horizons according to our choice of ensemble, we can verify the first law. To leading order in $T_+$, the mass $M$, entropy $S_+$ and chemical potential $\mu_+$ of the outer event horizon change in the following way:
\begin{equation}
    \begin{aligned}\label{eq:thermo-cold}
        M = ~& M_0 + \frac{T_+^2}{M_{\text{gap}}^{\text{cold}}} + \mathcal{O}\left( T_+^3\right)\, , \\
        S_+ = ~& S_{0}+\frac{2T_{+}}{M^{\text{cold}}_{\text{gap}}}  + \mathcal{O}\left(T_{+}^{2}\right)\,, \\
        \mu_+ = ~& \mu_0 - \frac{Q_0 T_+}{M^{\text{cold}}_{\text{gap}} r_0^3} + \mathcal{O}\left(T_{+}^{2}\right)\,,
    \end{aligned}
\end{equation}
and $Q=Q_0$. Here, we have defined the mass gap
\begin{align}
    M_{\text{gap}}^{\text{cold}}  = \frac{1}{2\pi \ell_{\text{AdS}}^2 r_0}\,,
\end{align}
and $S_0 = \pi r_0^2$ is the extremal entropy, $\mu_0=Q_0/r_0$, and $M_0$ is given in \eqref{EqColdQ0M01}. Indeed, to leading order in $T_+$, these quantities obey 
\begin{align}
    d(M-\mu_+ Q)= T_+ dS_+-Qd\mu_+ \,,
\end{align}
the first law in the canonical ensemble. 

With this, it is convenient to revisit the decoupling limit \eqref{eq:decouple-cold} to incorporate the leading order effect in $T_+$. For this reason, we trade the decoupling parameter $\lambda$ in \eqref{eq:decouple-cold} for $T_+$ and implement the decoupling limit by defining
\begin{align}
    r = r_+ + 4 \pi T_+ \ell_{\text{AdS}}^2 \sinh^2 \frac{\eta}{2}\,, \quad \qquad t = \frac{1}{2\pi T_+} \left( - i \tau \right)\,.
\label{EqDecouplingCold}
\end{align}
and then taking $T_+\to0$ on the black hole background. Here, $\eta \in \left[ 0, \infty \right)$ is a new radial coordinate which parametrizes our distance from the extremal horizon (located at $\eta = 0$). We also go to Euclidean signature, where the coordinate $\tau \in \left[ 0, 2 \pi \right]$ corresponds to a Wick-rotated time coordinate on a thermal circle of periodicity $T_+$. Using \eqref{Eqcolddecoupling} and \eqref{EqDecouplingCold} on \eqref{eq:4dvanillametric}-\eqref{eq:gaugefield}, to leading order in $T_+$ we obtain
\begin{equation}
    \begin{aligned} \label{geom_cold_exp}
    g_{\mu\nu} &= \bar{g}_{\mu\nu} +  T_+\,\delta g_{\mu\nu}\,   +  \mathcal{O} \left( T_+^2 \right)\,, \\
    A &= \bar{A} +  T_+ \,\delta A\,  + \mathcal{O} \left( T_+^2 \right)\,,
\end{aligned}
\end{equation}
where 
\begin{equation}\label{eq:ads2cold}
    \begin{aligned}
   \bar{g}_{\mu\nu} \d x^\mu \d x^\nu &=  \ell_{\text{AdS}}^2 \left( \sinh^2 \eta \d \tau^2 + \d \eta^2 \right) + r_0^2 d\Omega_2^2\,, \\
 \bar A &=  i Q_0\frac{\ell_{\text{AdS}}^2}{r_0^2}  \left( \cosh \eta - 1 \right)\d \tau\,,   
\end{aligned}
\end{equation}
is the expected EAdS$_2\times S^2$ near horizon geometry appearing at extremality; it is the Euclidean counterpart of  \eqref{EqNHColdGeometry}-\eqref{eq:gauge-cold} and they are related locally by a diffeomorphism. The first order response in $T_+$ in \eqref{geom_cold_exp} reads  
\begin{equation}
    \begin{aligned} \label{eq:firstT-cold}
 \frac{ \delta g_{\mu\nu} \d x^\mu \d x^\nu}{4\pi \ell_{\text{AdS}}^2} &=  
    \frac{\ell_{\text{AdS}}^2 }{3r_0^3} \left( \ell_{\text{AdS}}^2 + 2r_0^2 \right) \left( 2 + \cosh \eta \right) \tanh^2 \frac{\eta}{2} \left( - \sinh^2 \eta \d \tau^2 + \d \eta^2 \right) \\&\qquad + r_0 \cosh \eta d\Omega_2^2\,,\\
 \delta A &= - i  \frac{2\pi \ell_{\text{AdS}}^4 }{r_0^3} Q_0 \sinh^2 \eta \,\d\tau\,.
\end{aligned}
\end{equation}
The background \eqref{geom_cold_exp}-\eqref{eq:firstT-cold} will be the configuration we will use to evaluate the path integral along the lines of the discussion in Sec.\,\ref{sec:away}.  Note that whilst the leading order correction to the graviton operator is $\mathcal{O} \left( T_+ \right)$, it is necessary to work to $\mathcal{O} \left( T_+^2 \right)$ in \eqref{Eqcolddecoupling} due to the rescaling of time in the decoupling limit \eqref{EqDecouplingCold}.

In the above discussion, we have ignored that our black hole is embedded in dS$_4$; this adds a cosmological horizon at $r_c$ which carries its own finite temperature $T_c \gg T_+$. By taking the decoupling limit \eqref{EqDecouplingCold}, we have zoomed in locally to the black hole throat so that the cosmological horizon is infinitely far away, and an observer could not see it.\footnote{
In Sec.\,\ref{SecNariaiCorrections} we will again have to decide where we locate our observer.
} Our stand will be that \eqref{geom_cold_exp} is a solution to Einstein's equations locally in the throat, and we are computing the path integral there, which does not include the cosmological horizon at $r_c$.
If one were computing the path integral in the full spacetime geometry, as in \cite{Kapec:2024zdj,Kolanowski:2024zrq}, one would be concerned about how to treat the Euclidean geometry containing horizons with different temperatures. We will comment further on this in Sec.\,\ref{sec:conclusion}.

\subsection{Near-extremal path integral}\label{sec:zero-cold}

Having described the near horizon geometry of the Cold black hole, we now turn to the path integral around this background. First, we will report on each of the zero modes that appear in the extremal limit, along the lines of Sec.\,\ref{subsubtensor}, and then see how the eigenvalues of these modes are corrected by the near-extremal background \eqref{geom_cold_exp}, following the discussion in Sec.\,\ref{sec:away}. We note that several aspects of this section resemble the analysis in \cite{Iliesiu:2022onk,Banerjee:2023quv}, as those works also consider a near-extremal geometry starting from AdS$_2\times S^2$. The presence of the four-dimensional cosmological constant does introduce some new aspects related to vector and photon modes, which we will highlight below.

\paragraph{Tensor modes.}
As previewed in Sec.\,\ref{subsubtensor}, the first kind of zero modes we consider are graviton modes corresponding to normalisable perturbations of the EAdS$_2\times S^2$ metric, generated by a non-normalisable diffeomorphism $\zeta$. 
Following the analysis there, first we need to solve \eqref{EqTransversecond} on the AdS$_2$ background; using the coordinates in \eqref{eq:ads2cold}, we find
\begin{align}
    \Phi_n = e^{in\tau} \tanh^n \left( \frac{\eta}{2}\right) \frac{\ell_{\text{AdS}} \,  \left( |n| + \cosh \eta \right)}{\sqrt{32|n|\left(n^2 -1 \right)}\pi r_0 }\,,
\end{align}
where $n\in \mathbb{Z}$; note that the normalisation has been chosen poorly for  $n=0,\pm1$, but these configurations will be removed shortly. A key property of these solutions is that they are designed to be single-valued in $\tau\sim \tau+2\pi$, which will assure regularity and a well-behaved norm.  The corresponding vector, defined in \eqref{EqZetainPhi}, reads
\begin{equation}
\begin{aligned} \label{diffeos_AdS2}
    \zeta_n = & \frac{e^{in\tau}  \tanh^n \left( \eta/2\right) }{\sqrt{32 |n| \left( |n|^2 - 1 \right)}\pi r_0 \ell_{\text{AdS}}} \Bigg( \frac{|n|\left( |n| + \cosh \eta \right) + \sinh^2 \eta}{\sinh^2 \eta} \partial_{\tau} \\  & \qquad\qquad\qquad\qquad \qquad\qquad\qquad\qquad\qquad\qquad- \frac{i|n| \left( |n| + \cosh \eta \right)}{\sinh \eta} \partial_{\eta} \Bigg)\,,
\end{aligned}
\end{equation}
with $|n|\geq2$. The values $n=0,\pm 1$ are excluded here due to the fact that $\zeta_{-1}$, $\zeta_{0}$, and $\zeta_{1}$ are isometries of the EAdS$_2 \times S^2$ near horizon geometry and hence their Lie derivative vanishes. 
The metric perturbations generated by \eqref{diffeos_AdS2} are then
\begin{align} \label{metric_pert_cold}
     h^{(n)}_{\mu \nu} \d x^{\mu}\d x^{\nu}    = &  \frac{ie^{in\tau}\ell_{\text{AdS}} \sqrt{|n|(n^2-1)}}{\sqrt{8}\pi r_0} \tanh^n \left(\frac{\eta}{2} \right) \left( \frac{\d \eta^2}{\sinh^2 \eta} + \frac{2i\d \eta \d \tau}{\sinh \eta} - \d \tau^2 \right)\,,
\end{align}
where $|n|\geq 2$, and they are correctly normalised, i.e.,
\begin{align}
    \int d^4 x \sqrt{\bar{g}} \, {h^{(n)}_{\mu \nu}}^* {\bar{g}}^{\mu \alpha}{\bar{g}}^{\nu \beta} h^{(m)}_{\alpha \beta} = \delta^{mn}\,.
\end{align}
One can show that all the metric perturbations in  \eqref{metric_pert_cold} are annihilated by the operator $D$ from \eqref{EqLinGravOperator} and thereby are the desired tensor zero modes.
 Since there are infinitely many of these modes, they introduce a divergence in the extremal path integral as described in \eqref{eq:zm}.

As explained in Sec.\,\ref{sec:away}, we can regulate the divergence of these zero modes by introducing a correction to the extremal geometry.  Our corrected background is the near-extremal geometry \eqref{geom_cold_exp}-\eqref{eq:firstT-cold}, which includes a first-order effect due to turning on $T_+$ near the horizon. The relevant correction to the eigenvalue is given by \eqref{eq:correc-D}
\begin{align}
    (\delta \Lambda_n)_{\text{tensor}} =  \int d^4x  \sqrt{g}  \,{h^{(n)}_{\alpha \beta}}^* \delta D^{\alpha \beta \mu \nu} h^{(n)}_{\mu \nu}  = \frac{|n| T_+}{16r_0} + \mathcal{O} \left( T_+^2 \right)\,,
\label{EqAdShDh}
\end{align}
where we used \eqref{EqLinGravOperator} together with  \eqref{metric_pert_cold} and \eqref{geom_cold_exp} to first order in $T_+$. This makes it evident that the eigenvalue of a tensor mode is corrected by a finite temperature effect. Next, evaluating the corresponding path integral for these modes will take the form \eqref{eq:Zcorr}. The resulting expression is 
\begin{equation}
\begin{aligned}\label{eq:Ztensor-cold}
    \left( \delta \log Z \right)_{\text{tensor}} &= - \frac{1}{2} \sum_{\left| n \right| \geq 2}  \log \left( \delta \Lambda_n \right)_{\text{tensor}}\\
    &= - \log \prod_{n\geq 2} \left( \delta \Lambda_n \right)_{\text{tensor}} \\
    &= \frac{3}{2} \log \frac{T_+}{r_0} - \log \left( 64 \sqrt{2\pi} 
 \right) + \cdots\,,
\end{aligned}
\end{equation}
where we used \eqref{EqAdShDh} and the zeta function regularization 
\begin{equation}\label{eq:zeta2}
\prod_{n\geq 2} \frac{\alpha}{n} =  \frac{1}{\alpha^{3/2}\sqrt{2\pi}}\,,
\end{equation} 
with $\alpha$ a real parameter. The dots in \eqref{eq:Ztensor-cold} reflect that we are only reporting on the leading order effect in $T_+$.

It is instructive to briefly expand on the geometrical interpretation of these tensor modes and the associated diffeomorphisms \eqref{diffeos_AdS2}, which is common in analysing path integrals near to AdS$_2$ \cite{Charles:2019tiu,Iliesiu:2020qvm,Iliesiu:2022onk}. Redefining $\zeta = \sum_{n} \epsilon^{(n)} \zeta_n $, and introducing a new function $\epsilon(\tau) = \sum_n \epsilon^{(n)} e^{i n \tau} $, we find the vector field at the boundary of AdS$_2$ is
\begin{align}
    \begin{split}
        \lim_{\eta \rightarrow \infty} \zeta  \propto \epsilon(\tau) \partial_{\tau} - \epsilon'(\tau) \partial_{\eta}\,.
    \end{split}
\label{EqAdSVectors}
\end{align}
 Therefore, these diffeomorphisms correspond to a boundary time reparameterisation that sends $( \tau,\eta ) \rightarrow ( \tau - \epsilon(\tau), \eta + \epsilon'(\tau))$. The path integral then is proportional to an integral over the (infinite-dimensional, non-compact) coset $\textit{Diff}(S^1)/SL(2,\mathbb{R})$, where the quotient arises because the $n = 0, \pm 1$ perturbations are excluded. Regulating this path integral by turning on the temperature falls along the lines of the symmetry breaking pattern \cite{Maldacena:2016upp}, which is known to control many aspects of the physics of black holes near extremality.

\paragraph{Vector modes.}
\label{SecAdSVector}
We will now consider the contribution from the vector modes that arise by diffeomorphisms of the form \eqref{eq:diff-vec-modes} acting on the EAdS$_2\times S^2$ geometry in \eqref{eq:ads2cold}. These are determined by solutions to \eqref{EqLargeGauge}, which read
\begin{align}
    \Phi^0_n = \frac{1}{\sqrt{2\pi^3 |n|} r_0^2} \left( \frac{\sinh \eta}{1+\cosh \eta} \right)^n e^{in\tau}\,, 
\label{EqColdGaugeScalar}
\end{align}
where $|n|\geq 1$ and solutions are single-valued in $\tau\sim\tau +2\pi$; the mode with $n=0$ has been excluded since it does not give rise to a metric perturbation.  The vector modes are then given by \eqref{eq:vector-mode} 
\begin{align}
    \begin{split}
        h_{\mu \nu} \d x^{\mu} \d x^{\nu} = & \sum_{|n|\geq 1 } \sum_{m=-1,0,1} v^{(n,m)}_{\mu \nu} \d x^{\mu} \d x^{\nu} \\
        = & \sum_{|n|\geq 1 } \sum_{m=-1,0,1} \epsilon_{ij} \partial^{j} Y_{l=1,m}\left( \theta, \phi \right) \partial_{a} \Phi_n^0 \left( \tau, \eta \right) \d x^{i} \d x^{a}\,.
    \end{split}
\label{EqAdSVectorModes}
\end{align}
As with the tensor modes, these modes are normalised such that
\begin{align}
    \int d^4 x \sqrt{\bar{g}} \, {v^{(m,n)}_{\mu \nu}}^* {\bar{g}}^{\mu \alpha}{\bar{g}}^{\nu \beta} v^{(m',n')}_{\alpha \beta} = \delta^{mm'}\delta^{nn'}\,.
\end{align}
From the structure of the metric components in \eqref{EqAdSVectorModes}, it seems like one could build a vector mode by using $\epsilon_{ab} \partial^{b} \Phi^0$. However, the solutions in  \eqref{EqColdGaugeScalar} satisfy
\begin{align}
    i \epsilon_{ab} \partial^{b} \Phi^0_n = \partial_{a} \Phi^0_n\,.
\end{align}
Therefore, the dualized vectors do not produce additional zero modes. Also, the vector zero modes ${v^{(m,n)}_{\mu \nu}}$ are orthogonal to the tensor modes in \eqref{metric_pert_cold}, and therefore there is no mixing. To conclude, we have another infinite tower of modes, which annihilate $D$ in 
\eqref{EqLinGravOperator} and have a finite norm.

With this, we can now implement the analysis of Sec.\,\ref{sec:away} for our vector modes. Following the same steps as for the tensor modes, we find that for all three values of $m$, the vector modes \eqref{EqAdSVectorModes} give the same finite temperature contribution to the graviton operator. In concrete, we find
\begin{align}
    (\delta \Lambda_n)_{\text{vector}} =  \int d^4x \sqrt{g} \,{v^{(m,n)}_{\alpha \beta}}^* \delta D^{\alpha \beta \mu \nu} v^{(m,n)}_{\mu \nu}  = \frac{\left( \ell_{\text{AdS}}^2 + 2r_0^2 \right) |n|T_+}{48r_0^3} + \mathcal{O} \left( T_+^2 \right)\,.
\label{EqAdSvDv}
\end{align}
where we evaluated \eqref{eq:correc-D} using \eqref{geom_cold_exp} and \eqref{EqAdSVectorModes} to leading order in $T_+$.
For each value of $m$, we sum up these contributions as %
\begin{equation}
\begin{aligned}
    %
        -\frac{1}{2} \sum_{|n|\geq 1} \log(\delta \Lambda_n )_{\text{vector}} &=
    - \log \prod_{n\geq 1} (\delta \Lambda_n )_{\text{vector}}\\
    &= \frac{1}{2} \log \frac{\left( \ell_{\text{AdS}}^2 + 2r_0^2 \right) T_+}{r_0^3} - \log \left( 4\sqrt{6\pi} \right)\,,
\end{aligned}
\end{equation}
noting now that we used
\begin{equation}\label{eq:zeta1}
    \prod_{n\geq 1} \frac{\alpha}{n} =  \frac{1}{\sqrt{2\pi \alpha}}~.
\end{equation}
 The total vector contribution, considering each value of $m$, is thus
\begin{align}\label{eq:Zvector-cold}
    \left( \delta \log Z \right)_{\text{vector}} = \frac{3}{2} \log \frac{\left( \ell_{\text{AdS}}^2 + 2r_0^2 \right) T_+}{r_0^3} - 3 \log \left( 4\sqrt{6\pi} \right)+\cdots~.
\end{align}

\paragraph{Gauge (photon) modes.}\label{sec:coldphoton}
Finally, we report on the gauge (photon) modes described in Sec.\,\ref{subsubGauge}. For the Cold limit, we construct our photon modes from the scalars \eqref{EqColdGaugeScalar} generating large gauge transformations as
\begin{align}\label{eq:photon-cold}
    {a^{(n)}} = \frac{\sqrt{\pi} r_0}{2} \partial_{a}\Phi_n^0 \d x^a~.
\end{align}
These have been normalised such that
\begin{align}
    \int d^4 x \sqrt{\bar{g}} \, {a^{(n)}_{\mu}}^* {\bar{g}}^{\mu \nu} {a^{(m)}_{\nu}} = \delta^{mn}~.
\end{align}
Here, the index $n$ runs over $|n| \geq 1$. It is worth noting that $a^{(n)}$ is subleading at the AdS$_{2}$ boundary relative to the background field $\bar A$ in \eqref{eq:ads2cold}, i.e., in the limit $\eta \to\infty$ we have $a^{(n)}\sim O(1)$ whereas $\bar A\sim Q_0 e^{\eta}$. Hence the fluctuations will not change the electric charge measured in the throat.  The modes $a^{(n)}$ also annihilate the operator $P$ in \eqref{EqPhotonOperator}, and $O_{\text{int}}$ in \eqref{EqIntha} at extremality ($T_+=0$). For all of these reasons, the modes in \eqref{eq:photon-cold} are zero modes in our path integral.

With this, we compute the eigenvalue correction of these modes when considering corrections near to extremality as before. By using \eqref{eq:correc-P} and \eqref{geom_cold_exp}, we see that the corrected eigenvalue relative to $P$ is
\begin{align}
    (\delta \Lambda_n)_{\text{photon}} =  \int d^4 x \sqrt{g} {a_{\mu}^{(n)}} \delta P^{\mu \nu} a_{\nu}^{(n)} = \frac{\left( \ell_{\text{AdS}}^2 - r_0^2 \right) |n| T_+}{24r_0^3} + \mathcal{O} \left( T_+^2 \right).
\label{EqaDaCold}
\end{align}
Firstly, note that $\ell_{\text{AdS}}^2 > r_0^2$, so this contribution always has a definite positive sign. Secondly, observe that in the limit of zero bulk cosmological constant ($\Lambda = 0$) we have $\ell_{\text{AdS}}=r_0$, and the correction \eqref{EqaDaCold} vanishes. Thus, the linear in $T_+$ contribution from the photon modes vanishes for extremal black holes embedded in four-dimensional asymptotically flat space, in agreement with \cite{Banerjee:2023quv}. The result in \eqref{EqaDaCold} is then a new contribution from the gauge field zero mode, due to the presence of a cosmological constant. The photon zero modes also contribute in the presence of a negative bulk cosmological constant, namely in the case of near-extremal RN-AdS black hole, as shown in Sec.\,\ref{AdS4_subs}. 

It is worth mentioning the role of the mixing term $O_{\text{int}}$ in correcting the eigenvalues of our tensor, vector, and gauge zero modes. One can check explicitly that on the background \eqref{geom_cold_exp}, and for the zero modes constructed in this section, there is no contribution from \eqref{EqIntha} to leading order in $T_+$, i.e., $\langle h| \delta O_{\text{int}} a\rangle=0$. We have also not found additional zero modes that are intrinsic to $O_{\text{int}}$ itself. 


To conclude, the correction to the path integral due to the zero modes of photon  yields 
\begin{equation}\label{eq:Zphoton-cold}
\begin{aligned}
    \left( \delta \log Z \right)_{\text{photon}} &= -\frac{1}{2} \log \prod_{|n|\geq 1} (\delta \Lambda_n)_{\text{photon}}\\
    &= \frac{1}{2} \log \frac{\left( \ell_{\text{AdS}}^2 - r_0^2 \right) T_+}{r_0^3} - \log \left( 4 \sqrt{3\pi} \right)+\cdots\,,
\end{aligned}
\end{equation}
where we have used the same regularisation as in \eqref{eq:zeta1}.

\paragraph{Summing all zero mode contributions.}\label{sec:summingcold}
In this final portion, we collect the contribution of each zero mode (tensor, vector and gauge). This gives
\begin{equation}
\begin{aligned}
    \delta \log Z_{\text{z.m}} &= (\delta \log Z)_{\text{tensor}} + (\delta \log Z)_{\text{vector}} +(\delta \log Z)_{\text{photon}}\\
    &= \frac{3}{2} \log \frac{T_+}{r_0} + \frac{3}{2} \log \frac{\left( \ell_{\text{AdS}}^2 + 2r_0^2 \right) T_+}{r_0^3} + \frac{1}{2} \log \frac{\left( \ell_{\text{AdS}}^2 - r_0^2 \right) T_+}{r_0^3} + \cdots\,,
\end{aligned}
\end{equation}
where used \eqref{eq:Ztensor-cold}, \eqref{eq:Zvector-cold} and \eqref{eq:Zphoton-cold}. We have also discarded the constant terms which do not depend on any physical parameters. The low-temperature dependence of the path integral therefore scales as 
\begin{align}
    \log Z \sim \frac{7}{2} \log T_+~.
\label{EqAdSResult}
\end{align}
We note the difference in the result reported in \cite{Banerjee:2023quv}, where $\Lambda = 0$. This is due to the additional contribution from the photon modes, as discussed above. The result \eqref{EqAdSResult} will provide a useful comparison when we discuss the Nariai case.


%
\subsection{Corollary: the near-extremal RN black hole in \texorpdfstring{AdS$_{4}$ }{AdS4} }\label{AdS4_subs}

We can easily extend the results we have obtained so far to the case of extremal static black hole in AdS$_4$ (with $\Lambda<0$). Indeed the latter is related to the Cold limit of RNdS$_4$ by analytically continuing $\ell_4 \rightarrow i \ell_4$, which effectively switches the sign of the cosmological constant. In this case we have only one extremal ``cold'' limit, characterized by parameters \cite{Romans:1991nq}
\begin{align}
    Q_0^2 = r_0^2 \left( 1 + 3 \frac{r_0^2}{\ell_4^2} \right)~, \qquad M_0 = r_0 \left( 1 + 2 \frac{r_0^2}{\ell_4^2} \right)~. 
\label{AdS_Cold}
\end{align}
Working again in an ensemble of fixed charge, by taking the analytic continuation of \eqref{eq:Ztensor-cold}, \eqref{EqAdSvDv}, and \eqref{EqaDaCold}, the contributions from tensor modes, vector and gauge modes read respectively
\begin{equation}
 \begin{aligned}
 (\delta {\Lambda_n})_{\text{tensor,\,AdS}} &=  \frac{|n| T_+}{16r_0} + \mathcal{O} \left( T_+^2 \right)\,,  \quad \qquad \qquad \qquad \quad ~~\left| n \right| \geq 2\,,  \\
  (\delta {\Lambda_n})_{\text{vector,\,AdS} } &=  \frac{\left( \ell_{\text{AdS}}^2 + 2r_0^2 \right) \left| n \right| T_+ }{48r_0^3} + \mathcal{O} \left( T_+^2 \right)\,, \, \quad \qquad \, \left| n \right| \geq 1 \,,\\
  (\delta {\Lambda_n})_{\text{gauge,\,AdS} } &=  \frac{\left( \ell_{\text{AdS}}^2 - r_0^2 \right) \left| n \right| T_+}{24r_0^3}  + \mathcal{O} \left( T_+^2 \right), \quad \quad \qquad \left| n \right| \geq 1\,,
\label{AdS_Cold_corr}    
 \end{aligned}   
\end{equation}
where the effective AdS$_2$ length is now $\ell_{\text{AdS}}^2 = \ell_4^2 r_0^2 / \left( \ell_4^2 + 6r_0^2 \right)$. Summing together each contribution \eqref{AdS_Cold_corr} as before, accounting for all the $m \in \lbrace 0, \pm 1 \rbrace$ and completing the zeta function regularisation, we return the same total contribution
\begin{align}
    \log Z \sim \frac{7}{2} \log T_+\,.
\end{align}
Let us mention that the result for the tensor modes here agrees with the one found in \cite{Maulik:2024dwq} for extremal Kerr-Newman AdS$_4$ black holes in the limit in which the angular momentum vanishes. The results for the vector modes and gauge modes are instead genuinely new.

\section{Path integral for the near-Nariai limit of \texorpdfstring{SdS$_4$}{SdS4}}
\label{SecSdSNariai}
When studying the Nariai limit of the RNdS$_4$ black hole, we will discover a number of issues with the behaviour of zero modes in the extremal and near-extremal limit, that ultimately cause problems in the Euclidean path integral. To decode the origin of this issue, we will begin by considering the chargeless case, that is, the Nariai limit of the Schwarzschild-de Sitter geometry (SdS$_4$). This will demonstrate that the problems we uncover are due to properties of the Euclidean geometry and are not caused by the presence of gauge fields. 

The SdS$_4$ case provides a natural point of comparison to the works in  \cite{Maldacena:2019cbz,Cotler:2019nbi,Moitra:2022glw}. There, one-loop contributions to the \textit{Lorentzian} wave function were calculated using the boundary Schwarzian action characteristic of 2D JT gravity. As those computations were completed from a 2D perspective, they were only able to consider contributions to the wave function from the tensor modes. Thus, in what follows, the results we report for vector modes are new.\footnote{See also \cite{Turiaci:2025xwi} which revisits several aspects of \cite{Maldacena:2019cbz} to incorporate appropriately the four-dimensional physics when the boundary of dS$_4$ is $S^1\times S^2$.} In short, here we will take a Euclidean approach to implement the analysis of zero modes in Sec.\,\ref{Sec3GenStrat}, and work directly in four dimensions, which will highlight several subtleties that are not present in JT gravity.\footnote{One can modify the effective theory in 2D to include the massless fields that are responsible for this. This has not been done in the literature and we will not pursue this route here. }

In this section, we will begin by discussing the Nariai limit of the SdS$_4$ geometry introduced in Sec.\,\ref{SecRNdSandSdS}. As discussed there, the near-horizon geometry of the Nariai limit is $\text{dS}_2 \times S^2$. In the Nariai limit of SdS$_4$, the only corrections to the path integral are due to the graviton operator, which means that we only need to consider the tensor and vector modes. We will begin in Sec.\,\ref{SubSecHorSep} by perturbing the Nariai black hole following a similar procedure as in the Cold case in Sec.\,\ref{SecColdLimit}, by heating up the extremal horizon to a finite temperature and introducing a separation between the horizons. We will discover an issue with the corrected eigenvalues of modes, and the norms of the vector modes, which means that we cannot define a convergent path integral unless we analytically continue them in an ad-hoc way. We will also consider in Sec.\,\ref{SubSecHorizoncoinc} a different deformation of the $\text{dS}_2 \times S^2$ geometry, which is to keep the horizons coincident at zero temperature and explore the effects of moving outwards the throat. From a 2D perspective, this is equivalent to having a running dilaton. We will find the same problems with convergence of the path integral in that case.

\subsection{Finite temperature response of the path integral}
\label{SubSecHorSep}
We heat up the Nariai black hole following a similar procedure as in the Cold black hole. In this case, the coincident horizons are $r_+$ and $r_c$, and in order to be able to successfully identify the asymptotic symmetries, we work in the Milne patch (see Fig.\,\ref{fig:penrosernds}) and place our observer outside the cosmological horizon. We deviate from extremality by increasing $T_c$, which causes $r_+$ to decrease and $r_c$ to increase, whilst $\ell_4$ remains fixed. The horizons shift as 
\begin{equation}
\begin{aligned}
    r_+ &= \frac{\ell_4}{\sqrt{3}}  - \frac{2\pi\ell_4^2}{3} T_c - \frac{10  \pi^2 \ell_4^3}{9\sqrt{3}} T_c^2 + \mathcal{O}\left(T_c^3\right) \,,\\
       r_c &= \frac{\ell_4}{\sqrt{3}}  + \frac{2\pi\ell_4^2}{3} T_c + \frac{2  \pi^2 \ell_4^3}{3\sqrt{3}} T_c^2 + \mathcal{O}\left(T_c^3\right) \,,
\end{aligned}
\end{equation}
where we used \eqref{EqThmuh}, \eqref{Eqrcinrp} and \eqref{EqNariaiSchw}. Although we are only interested in the first-order temperature correction to the Nariai background, as we will be performing a rescaling of time of order $T_{c}$, it is necessary to keep track of the change in $r_{+,c}$ to second order.  The response in the mass and entropy reads
\begin{equation}
    \begin{aligned}\label{eq:near-thermo-N}
        M&=M_N -\frac{T_c^2}{M_{\text{gap}}^{\text{Nariai}}} + \mathcal{O}\left( T_c^3 \right)~, \\
        S_c &=S_N +\frac{2T_c}{M_{\text{gap}}^{\text{Nariai}}} + \mathcal{O}\left( T_c^2 \right) ~,
    \end{aligned}
\end{equation}
where $M_N=r_{N}/3$ and $S_N=\pi r_{N}^2$, with $r_{N}=\ell_{4}/\sqrt{3}$ as defined in \eqref{EqNariaiSchw}. The mass gap is 
\begin{equation}
 M_{\text{gap}}^{\text{Nariai}}=\frac{1}{2\pi^2 r_N^3}~.
\end{equation}
As we deviate away from extremality, we have a negative heat capacity $C= \frac{\partial M}{\partial T}$. This is very different from the response of the Cold black hole in \eqref{eq:thermo-cold} which has a positive heat capacity. We have strong reasons to believe that this is the culprit of the peculiarities that will come.\footnote{We thank Z. Yang for bringing this to our attention.} 

In this portion, we will be deforming the extremal background \eqref{EqNariaiGeometry1}, with $Q_N=0$, by turning on $T_c$. The heated-up extremal geometry, which defines near-extremal Nariai, is obtained by introducing a decoupling limit in the original SdS$_4$ black hole controlled by $T_c$. Concretely we have
\begin{align}
    r = r_c + \frac{4\pi \ell_4^2}{3} T_c \sinh^2 \left( \frac{\eta}{2} \right)\,, \qquad t = \frac{1}{2\pi T_c} \left( - i \tau \right)\,,
\label{EqDecouplingNariaSdS}
\end{align}
where again $\eta \geq 0$ is again a coordinate parameterising our distance outside the horizon, and we have rescaled and Wick rotated the time coordinate. 
In this decoupling limit, $r \geq r_c$, so these coordinates cover the Milne patch (also known as cosmological interior), see Fig.\,\ref{fig:penroseSdS}.\footnote{Had we used $r \leq r_c$ and worked inside the static patch, upon Wick rotation our geometry would be $S^2 \times S^2$, a compact geometry. Thus, there would be no zero modes, and our computational techniques would not be applicable.}
Under the decoupling limit in \eqref{EqDecouplingNariaSdS}, the metric becomes
\begin{align} \label{geom_SdS_exp}
    g_{ab} = \bar{g}_{ab} + T_c\,\delta g_{ab} + \mathcal{O}\left(T_c^2\right)\,,
\end{align}
where
\begin{equation}
    \begin{aligned}\label{geom_SdS_exp-1}
    \bar{g}_{\mu \nu} \d x^{\mu} \d x^{\nu} = & -\frac{\ell_4^2}{3} \left( \sinh^2 \eta \d \tau^2 + \d \eta^2 \right) + \frac{\ell_4^2}{3} d\Omega_2^2\,, \\
    \delta g_{\mu \nu} \d x^{\mu} \d x^{\nu} = & -\frac{4 \pi \ell_4^3 }{9\sqrt{3}} \left( 2 + \cosh \eta \right) \tanh^2 \frac{\eta}{2}\left( - \sinh^2 \eta \d \tau^2 + \d \eta^2 \right) \\ &\quad + \frac{4 \pi \ell_4^3}{3\sqrt{3}} \cosh \eta  d\Omega_2^2\,.
\end{aligned}
\end{equation}
Here the coordinates cover $\eta \in \left[ 0, \infty \right)$ and $\tau \in \left[ 0, 2\pi \right]$. The geometry of $\bar g_{\mu \nu}$ is  $(-\text{EAdS}_2) \times S^2$. 
The minus sign in front of Euclidean AdS$_2$ will play an important role in the computations which follow. Also, we note that these results have the same form as in the Cold limit \eqref{geom_cold_exp}, with the replacements 
\begin{equation}\label{eq:cold-to-N}
 \ell_{\text{AdS}}^2 \rightarrow -\ell^{2}_{\text{dS}}=-\frac{\ell_4^2}{3}\,, \quad 
  r_0 \rightarrow r_{N}=\frac{\ell_4}{\sqrt{3}}\,, \quad
 T_+ \rightarrow -T_c\,.    
\end{equation}

We now proceed to construct the zero modes for the Nariai background, and evaluating the near-extremal path integral following the steps in Sec.\,\ref{Sec3GenStrat}. Given the similarities with the Cold limit as mentioned above, many of our results will closely resemble the derivations in Sec.\,\ref{sec:zero-cold}.

\paragraph{Tensor modes.}
We start by identifying and constructing the appropriate zero modes that act on $-\text{EAdS}_2$  portion of $\bar g_{\mu \nu}$ in \eqref{geom_SdS_exp-1}. The solutions to \eqref{EqTransversecond} in this case are
 \begin{align}
    \Phi_n = e^{in\tau} \tanh^n \left( \frac{\eta}{2}\right) \frac{\left( |n| + \cosh \eta \right)}{\sqrt{32|n|\left(n^2 -1 \right)}\pi }\,,
\end{align}
labelled by an integer $n$. These lead to the vector fields \eqref{EqZetainPhi} 
 \begin{equation}
 \begin{aligned}
 \label{EqvecSdS}
    \zeta_n = & \frac{3 e^{in\tau}  \tanh^n \left( \eta/2\right) }{\sqrt{32 |n| \left( |n|^2 - 1 \right)}\pi \ell_4^2} \Bigg(  \frac{|n|\left( |n| + \cosh \eta \right) + \sinh^2 \eta}{\sinh^2 \eta} \partial_{\tau}\\ &\qquad \qquad \qquad \qquad \qquad\qquad \qquad\qquad \qquad-i \frac{|n| \left( |n| + \cosh \eta \right)}{\sinh \eta} \partial_{\eta} \Bigg)\,,
\end{aligned}
\end{equation}
where $|n| \geq 2$. Notice that the values $n=0,\pm 1$ are discarded since they correspond to Killing isometries of the background metric. 
Acting with \eqref{EqvecSdS} on the background metric $\bar g_{ab}$ then leads to the metric fluctuations
\begin{align} \label{metric_pert_SdS}
     h^{(n)}_{\mu \nu} \d x^{\mu}\d x^{\nu}  = &  \frac{ie^{in\tau} \sqrt{|n|(n^2-1)}}{\sqrt{8}\pi} \tanh^n \left(\frac{\eta}{2} \right) \left( \frac{\d \eta^2}{\sinh^2 \eta} + \frac{2i\d \eta \d \tau}{\sinh \eta} - \d \tau^2 \right)\,,
\end{align}
which obey 
\begin{align}
    \int d^4 x \sqrt{\bar{g}} \, {h^{(n)}_{\mu \nu}}^* {\bar{g}}^{\mu \alpha}{\bar{g}}^{\nu \beta} h^{(m)}_{\alpha \beta} = \delta^{nm}\,.
\end{align}
Hence we have constructed an infinite tower of modes that will have zero eigenvalues with respect to $D$ in \eqref{EqLinGravOperator}. These are zero modes of the Nariai path integral in the Milne patch, and render an infinite contribution.

Next, to regulate this divergence we deform the background geometry as in \eqref{geom_SdS_exp}-\eqref{geom_SdS_exp-1}. Using \eqref{eq:correc-D} and \eqref{metric_pert_SdS}, we find
\begin{align}
    (\delta \Lambda_n)_{\text{tensor}} =   \int d^4x  \sqrt{g}  \,{h^{(n)}_{\alpha \beta}}^* \delta D^{\alpha \beta \mu \nu} h^{(n)}_{\mu \nu} = -\frac{|n| T_c}{16 r_N} + \mathcal{O} \left( T_c^2 \right).
\label{EqSdShDhCoinc}
\end{align}
The correction to eigenvalues is negative, and hence we do 
\textit{not} have a convergent Gaussian in our path integral. This sign can be traced back to a relative minus sign difference between the $-\text{EAdS}_2$ and $S_2$, which introduces relative minus signs in \eqref{geom_SdS_exp-1}.

\paragraph{Vector modes.}
To identify the vector modes, we first construct solutions to \eqref{EqLargeGauge} around the background $\bar g_{\mu \nu}$ in \eqref{geom_SdS_exp}. The result reads
\begin{align}
    \Phi_n^0 = \frac{\ell_4^2}{\sqrt{72\pi |n|} } \left( \frac{\sinh \eta}{1+\cosh \eta} \right)^n e^{in\tau}\,,
\label{EqdSGaugeScalar}
\end{align}
where as before $\left| n \right| \geq 1$ and $\tau \sim \tau + 2\pi$. We construct vector modes using \eqref{eq:vector-mode} and the result takes the same form as in the Cold limit reported in \eqref{EqAdSVectorModes}, with appropriate changes in \eqref{eq:cold-to-N}. Notably, the normalisation of these modes is now \textit{negative}; that is, 
\begin{align}
    \int d^4 x \sqrt{\bar{g}} \, {v^{(m,n)}_{\mu \nu}}^* {\bar{g}}^{\mu \alpha}{\bar{g}}^{\nu \beta} v^{(m,n)}_{\alpha \beta} = -\delta^{nn'}\delta^{mm'}.
\label{EqSdSvecnorm}
\end{align}
This negative sign is again due to the relative minus sign between the $-\text{EAdS}_2$ and $S_2$ parts of the Euclidean metric. One ``leg'' of the vector mode $v_{\mu \nu}$ comes from the $-\text{EAdS}_2$ (the $\partial_{\mu} \Phi$) and the other ``leg'' from the $S_2$ (the $\epsilon_{\alpha \beta} \partial^{\beta} Y_{l=1,m}$), leading to an overall minus factor in the norm. By comparison, in the tensor mode \eqref{metric_pert_SdS}, both ``legs'' live in the $-\text{EAdS}_2$ part, the minus signs multiply together and the norm becomes one. The vector modes $v_{\mu \nu}$ do annihilate $D$ in \eqref{EqLinGravOperator}, and are orthogonal to the tensor modes \eqref{metric_pert_SdS}. Still their negative norm should raise a flag if they are truly zero modes in the path integral. Nevertheless, we will proceed and report on their behaviour in the near-extremal background and continue to refer to them as vector modes.

Following the same steps as before, we evaluate \eqref{eq:correc-D} for these negative norm states. We find 
\begin{align}
    \begin{split}
         \int d^4x \sqrt{g} \,{v^{(m,n)}_{\alpha \beta}}^* \delta D^{\alpha \beta \mu \nu} v^{(m,n)}_{\mu \nu}  = \frac{|n| T_c}{48r_N} + \mathcal{O} \left( T_c^2 \right)\,.
    \end{split}
\label{EqSdSvDv}
\end{align}
where all three values of $m=0,\pm1$ give the same value. Hence 
\begin{equation}
     (\delta \Lambda_n)_{\text{vector}}  = \frac{1}{\kappa_n}\langle h^{(n)} | \delta D h^{(n)}\rangle <0 ~, 
\end{equation}
since $\kappa_n=-1$ for the vector modes. The pathology here resembles the issue with the tensor modes since minus signs are appearing in unwanted locations. 

Encountering negative eigenvalues in a gravitational path integral is not new, with one standard reference being the negativity of conformal mode which unbounds effective action
\cite{Gibbons:1978ac}. The standard approach to this issue is to simply change the contour of integration in our path integral, which gives a finite answer albeit usually complex \cite{Polchinski:1988ua}.  Taking this approach we would conclude that the negative result in \eqref{EqSdShDhCoinc} should simply be rotated into the complex space and declare that the response of the path integral for the tensor modes at low temperatures is
\begin{equation}
    (\delta \log Z)_{\text{tensor}} \sim \log T_c^{3/2}~.
\end{equation}
This is the approach advocated in several other places, such as \cite{Gross:1982cv,Prestidge:1999uq}. A common thread in those references is that the appearance of the negative eigenvalue was tied to a thermodynamic instability. In our case, we have such instability, reflected in the response of the mass with temperature \eqref{eq:near-thermo-N}, making it also plausible here as the physical origin of such pathology in the Euclidean path integral. 
Encountering negative norm states is more unsettling. Still, this is not the first gravitational path integral to encounter such dragons. For instance, \cite{Marolf:2022ntb,Liu:2023jvm,Kolanowski:2024zrq} are prior works that encountered these obstacles as well. For the moment we will be agnostic about how to resolve this issue and explore how it manifests in the (near)-extremal limits that arise in the next section.

\subsection{Path integral near the throat of Nariai}
\label{SubSecHorizoncoinc}

In this portion, we will move away from the Nariai throat by allowing the geometry to expand outwards as an inflationary cosmology, without heating the system. Concretely, we will define the deformed geometry by introducing in SdS$_4$ the decoupling limit 
\begin{align}
    r = r_c + \frac{2\pi \ell_4^2}{3} \lambda R\,, \qquad t = \frac{1}{2\pi \lambda} \left(- i \tau \right)\,,
\end{align}
where we have Wick rotated $t$ to obtain a Euclidean geometry. Taking $\lambda\to 0$, the SdS$_4$ metric becomes
\begin{align}
    g_{\mu \nu} = \bar{g}_{\mu \nu} + \delta g_{\mu \nu} \lambda + \mathcal{O}(\lambda^2)~,
\end{align}
with
\begin{equation}
    \begin{aligned}\label{eq:metric-NC}
    \bar{g}_{\mu \nu} \d x^{\mu} \d x^{\nu} = & -\frac{\ell_4^2}{3} \left( R^2 \d \tau^2 + \frac{1}{R^2} \d R^2 \right) + \frac{\ell_4^2}{3} d\Omega_2^2\,, \\
    \delta g_{\mu \nu} \d x^{\mu} \d x^{\nu} = & -\frac{4 \pi \ell_4^3 R}{9\sqrt{3}} \left( -R^2 \d \tau^2 + \frac{1}{R^2} \d R^2 \right)+ \frac{4 \pi \ell_4^3}{3\sqrt{3}} R\,  d\Omega_2^2\,.
\end{aligned}
\end{equation}
Here $R \in [ 0, \infty )$ is a coordinate parameterising our distance outside the horizon that lies in the Milne patch of Fig.\,\ref{fig:penroseSdS}. 
In relation to JT gravity, what we have here is a nearly-dS$_2$ cosmology where the dilaton is expanding at a rate controlled by $\lambda R$ as $R\to \infty$. Our aim is to see how the Euclidean four-dimensional path integral responds to this class of background. We will focus on the zero modes appearing at $\lambda=0$ and how they are regulated when $\lambda$ is turned on. The results here should be compared with those in \cite{Turiaci:2025xwi}, where they also find a set of zero modes with large quantum effects.

\paragraph{Tensor modes.}
In a similar fashion as previous cases, we start by solving \eqref{EqTransversecond} on the background metric $\bar g_{\mu \nu}$ \eqref{eq:metric-NC};  this gives
\begin{align}
    \Phi_n = e^{in\tau-|n|/R} \frac{|n|+R}{\sqrt{32 |n|^3}\pi}~,
\end{align}
where $n \in \mathbb{Z}$ and $n = 0, \pm 1$ excluded.
The resulting vector fields are 
\begin{align}
    \zeta_n = \frac{3 e^{in\tau- |n|/R}}{\sqrt{32|n|^3}\pi \ell_4^2} \left( \frac{n^2 +|n| R + R^2}{R^2} \partial_{\tau} - i |n| \left( |n| + R \right) \partial_R \right)~,
\end{align}
which then generate the modes
\begin{align}
    h^{(n)}_{\mu \nu} \d x^{\mu} \d x^{\nu} = \frac{e^{in\tau- |n|/R}\sqrt{|n|^3}}{\sqrt{8}\pi} \left( \frac{\d R^2}{R^4} + \frac{2i \text{sign}(n) }{R^2} \d R\d \tau- \d \tau^2 \right)~,
\end{align}
and obey the normalisation
\begin{align}
    \int d^4 x \sqrt{\bar{g}} \, {h^{(n)}_{\mu \nu}}^* {\bar{g}}^{\mu \alpha}{\bar{g}}^{\nu \beta} h^{(m)}_{\alpha \beta} = \delta^{nm}~.
\end{align}
These modes will annihilate $D$ in \eqref{EqLinGravOperator} and hence correspond to zero modes in the path integral.  Next, by considering the deformation in \eqref{eq:metric-NC}, we can find the correction to the eigenvalues of these modes. The result is 
\begin{align}
    (\delta \Lambda_n)_{\text{tensor}} =   \int d^4x  \sqrt{g}  \,{h^{(n)}_{\alpha \beta}}^* \delta D^{\alpha \beta \mu \nu} h^{(n)}_{\mu \nu} = -\frac{|n| \lambda}{16 r_N} + \mathcal{O} \left( \lambda^2 \right)~.
\label{EqSdShDhcoinc}
\end{align}
We once again find that the tensor modes have a negative sign in their contribution to the graviton operator.

\paragraph{Vector modes.} The second class of modes are those that deform the sphere in \eqref{eq:metric-NC}. To construct them we first solve \eqref{EqLargeGauge} on $\bar g_{\mu \nu}$ which gives
\begin{align}
\Phi_n^0 = e^{in\tau - |n|/R} \frac{\ell_4^2}{\sqrt{72|n|\pi}}~,
\end{align}
restricting to $n \in \mathbb{Z}$ and $n \neq 0$.
The corresponding modes $v^{(m,n)}_{\mu\nu}$ are of the form \eqref{eq:vector-mode}. The norm of these states is
\begin{align}
    \int d^4 x \sqrt{\bar{g}} \, {v^{(m,n)}_{\mu \nu}}^* {\bar{g}}^{\mu \alpha}{\bar{g}}^{\nu \beta} v^{(m',n')}_{\alpha \beta} = -\delta^{nn'}\delta^{mm'}~,
\end{align}
which follows suit with \eqref{EqSdSvecnorm}.
Evaluating \eqref{eq:correc-D} for these modes gives
 \begin{align}
    \begin{split}
        \int d^4x \sqrt{g} \,{v^{(m,n)}_{\alpha \beta}}^* \delta D^{\alpha \beta \mu \nu} v^{(m,n)}_{\mu \nu} = \frac{|n| \lambda}{48r_N} + \mathcal{O} \left( \lambda^2 \right).
    \end{split}
\label{EqSdSvDvcoinc}
\end{align}
Not surprisingly, we obtain the same pathology as the near-extremal limit of Nariai case in \eqref{EqSdSvDv}.

The result in \eqref{EqSdShDhcoinc} and \eqref{EqSdSvDvcoinc} can be easily recovered by replacing  $T_c \rightarrow \lambda$ in \eqref{EqSdShDhCoinc} and \eqref{EqSdSvDv}, respectively. However, since we are now running the dilaton and working at zero temperature, we cannot attribute these pathologies to a thermodynamic instability. The analysis of \cite{Turiaci:2025xwi}, which considers the whole SdS$_4$ geometry, also finds non-positive eigenvalues in the one-loop corrections, however, the negativity of vector modes we encounter is not evident there. This provides some optimism that the pathologies we obtain are an artefact of the near horizon limit, but a more careful inspection would still be needed.

\section{Path integral for the near-Nariai limit of \texorpdfstring{RNdS$_4$}{}}
\label{SecNariaiCorrections}
Having analysed the neutral Nariai case, we now move on to the case where the charge is non-vanishing, $Q \not = 0$. 
We will discuss in particular different Euclidean incarnations of the near horizon solution. This, in essence, boils down to Sec.\,\ref{sec:decoup61} where we consider the geometry $(-\text{EAdS}_2 )\times S^2$, whereas in Sec.\,\ref{SubSecWickRot} we consider the geometry $-(\text{EAdS}_2 \times S^2)$. We will point out differences in the two cases for what concerns quantum contributions from zero modes to the path integral.
\subsection{Near-Nariai background: \texorpdfstring{$(-\text{EAdS}_2) \times S^2$}{}}\label{sec:decoup61}
We heat up the charged Nariai black hole following a similar procedure as in the previous sections. As in Sec.\,\ref{SubSecHorSep}, we work in the Milne patch (see Fig.\,\ref{fig:penrosernds}), and deviate from extremality by slightly increasing $T_c$. 
As in the Cold case, we work in the canonical ensemble by fixing $Q$ to its extremal value in \eqref{EqNariaiQnMn} and require that $\ell_4$ is unchanged; this will determine the response of all the parameters as a function of $T_c$. The outer and cosmological horizons then change as
\begin{equation}
    \begin{aligned}
        r_+ = ~& r_N - 2\pi \ell_{\text{dS}}^2 T_c + \frac{2\pi^2 \ell_{\text{dS}}^4 \left( 2\ell_{\text{dS}}^2 - 7r_N^2 \right)}{3r_N^3} T_c^2 + \mathcal{O}\left( T_c^3 \right) \, , \\
        r_c = ~& r_N + 2\pi \ell_{\text{dS}}^2 T_c - \frac{2\pi^2 \ell_{\text{dS}}^4 \left( 2\ell_{\text{dS}}^2 - 3r_N^2 \right)}{3r_N^3} T_c^2 + \mathcal{O}\left( T_c^3 \right) \, ,
    \end{aligned}
\end{equation}
where the dS$_2$ radius is defined in \eqref{dS2_radius}. The thermodynamic quantities respond as
\begin{equation}
    \begin{aligned}
        M = ~& M_N - \frac{T_c^2}{M_{\text{gap}}^{\text{Nariai}}} + \mathcal{O}\left( T_c^3 \right), \\
        S_c = ~& S_N + \frac{2T_c}{M_{\text{gap}}^{\text{Nariai}}} + \mathcal{O}\left(T_c^2\right),\\
        \mu_c = ~& \mu_N - \frac{ Q_N T_c}{M_{\text{gap}}^{\text{Nariai}} r_N^3} + \mathcal{O}\left( T_c^2 \right)
        \,,
    \end{aligned}
    \label{thermo-rels-Nariai1}
\end{equation}
and $Q = Q_N$. We refer to \eqref{EqNariaiQnMnv2} for the values of $M_{N}$ and $Q_{N}$, and we have defined
\begin{align}\label{M-gap-Nariai}
M_{\text{gap}}^{\text{Nariai}} = \frac{1}{2\pi^2 \ell^2_{\text{dS}}r_N}\,,
\end{align}
$S_N = \pi r_N^2$ is the extremal entropy, and $\mu_N = Q_N/r_N$. These quantities again obey the first law in the canonical ensemble up to leading order in $T_c$. As in Sec. \ref{SubSecHorSep}, we have a negative heat capacity (at fixed charge).

We obtain the geometry near the hot extremal horizon $r_c$ by revisiting the decoupling limit introduced in Sec.\,\ref{sec:Nariai}. Now using $T_c$ as our decoupling parameter, we define
\begin{align}\label{eq:newcoords1}
    r = r_c + 4\pi T_c \ell_{\text{dS}}^2 \sinh^2 \frac{\eta}{2}\,, \qquad t = \frac{1}{2\pi T_c} \left( - i \tau \right)\,,
\end{align}
where $\eta \in \left[ 0 , \infty \right)$ is again a coordinate parameterising our distance outside the extremal horizon. We move to Euclidean signature by introducing the Wick-rotated time coordinate $\tau \in \left[ 0 , 2 \pi \right]$ on a thermal circle of periodicity $T_c$. The metric and gauge field become 
\begin{equation}
    \begin{aligned} \label{geom_Nariai_exp}
    g_{\mu \nu} & = \bar{g}_{\mu \nu} + T_c \,  \delta g_{\mu \nu}  +  \mathcal{O} \left( T_c^2 \right)\,,  \\
    A &= \bar{A} +  T_c \,\delta A\,  + \mathcal{O} \left( T_c^2 \right)\,,
\end{aligned}
\end{equation}
where
\begin{equation}
    \begin{aligned}\label{EqdSgeometry} 
   \bar{g}_{\mu\nu} \d x^\mu \d x^\nu &=  - \ell_{\text{dS}}^2 \left( \sinh^2 \eta \d \tau^2 + \d \eta^2 \right)  + r_N^2 d\Omega_2^2 \,, \\
 \bar A &=  i Q_N \frac{ \ell_{\text{dS}}^2}{r_N^2} \left( \cosh \eta -1 \right) \, \d\tau\,,   
\end{aligned}
\end{equation}
and
\begin{equation}
    \begin{aligned}\label{eq:RN-Nariai-near}
 \frac{ \delta g_{\mu\nu} \d x^\mu \d x^\nu}{4\pi \ell_{\text{dS}}^2} &=  \frac{\ell_{\text{dS}}^2  }{3r_N^3} \left( \ell_{\text{dS}}^2-2r_N^2 \right)\left( 2 + \cosh \eta \right)\tanh^2 \frac{\eta}{2}\left( - \sinh^2 \eta\, \d \tau^2 + \d \eta^2  \right) \\&\qquad+ r_N \cosh \eta\, d\Omega_2^2 \,,\\
 \delta A &= - i 2\pi \frac{ \ell_{\text{dS}}^4}{r_N^3} Q_N \sinh^2 \eta \,\d\tau\,.
\end{aligned}
\end{equation}
 As in Sec.\,\ref{SecSdSNariai}, the near horizon Lorentzian geometry is dS$_2 \times S^2$, where we have again Wick rotated to $(-\text{EAdS}_2) \times S^2$. 

Note that the coefficient of $\delta g_{\tau\tau}$ and $\delta g_{\eta\eta}$ is related to the temperature difference between the coincident horizons, i.e., $\left( T_c - T_+ \right) \propto \left( \ell_{\text{dS}}^2 - 2 r_{N}^2 \right) T_c^2$. When $\ell_{\text{dS}}^2 = 2 r_{N}^2$ we have that $Q_N = M_N$:  the Nariai black hole becomes lukewarm. Therefore the sign of the metric perturbation in that sector flips at the lukewarm point. In particular, at the lukewarm point, only the $S^2$ is perturbed by the heating up of the geometry.

We emphasize that all of the results in the Nariai limit sofar have the same form as in the Cold limit, with the replacements 
\begin{equation} \label{eq:map-cold-N}
 \ell_{A}^2 \rightarrow -\ell_{\text{dS}}^2\,,\quad r_0 \rightarrow r_N\,,\quad T_+ \rightarrow - T_c \,.    
\end{equation}
This observation is analogous to \eqref{eq:cold-to-N}. For this reason, for the tensor and vector modes, we will again find the same problems as in Sec.\,\ref{SubSecHorSep}; the new twist is an additional issue arising from the photon contribution. In  Sec.\,\ref{SubSecWickRot} we will be able to provide a partial resolution to these undesirable features.
\paragraph{Tensor modes.}
\label{SecdSTensor}
As in Sec.\,\ref{sec:zero-cold} and Sec.\,\ref{SubSecHorSep}, we start by characterizing the so-called tensor modes. The relevant extremal geometry is $(-\text{EAdS}_2) \times S^2$ given in \eqref{EqdSgeometry}. Solving \eqref{EqTransversecond}, this time gives
\begin{align}
    \Phi_n = e^{in\tau} \tanh^n \left( \frac{\eta}{2}\right) \frac{i \ell_{\text{dS}} \,  \left( |n| + \cosh \eta \right)}{\sqrt{32|n|\left(n^2 -1 \right)}\pi r_N }\,,
\end{align}
where $n \in \mathbb{Z}$ and $n = 0, \pm 1$ excluded. The resulting vector field is
%
\begin{equation}
\begin{aligned} \label{diffeos_dS2}
    \zeta_n = & \frac{i e^{in\tau}  \tanh^n \left( \eta/2\right) }{\sqrt{32 |n| \left( |n|^2 - 1 \right)}\pi r_N \ell_{\text{dS}}} \Bigg( \frac{|n|\left( |n| + \cosh \eta \right) + \sinh^2 \eta}{\sinh^2 \eta} \partial_{\tau} \\  & \qquad\qquad\qquad\qquad \qquad\qquad\qquad\qquad\qquad\qquad- \frac{i|n| \left( |n| + \cosh \eta \right)}{\sinh \eta} \partial_{\eta} \Bigg)\,,
\end{aligned}
\end{equation}
with $\left| n \right| \geq 2$. Again, the $n = 0, \pm 1$ values are excluded as the $\zeta_{-1}, \zeta_0, \zeta_1$ are isometries of the near horizon geometry and their Lie derivative vanish. The metric perturbations generated by \eqref{diffeos_dS2} are 
\begin{align} \label{metric_pert_Nariai}
     h^{(n)}_{\mu \nu} \d x^{\mu}\d x^{\nu}    = &  \frac{e^{in\tau}\ell_{\text{dS}} \sqrt{|n|(n^2-1)}}{\sqrt{8}\pi r_N} \tanh^n \frac{\eta}{2} \left( \frac{\d \eta^2}{\sinh^2 \eta} + \frac{2i\d \eta \d \tau}{\sinh \eta} - \d \tau^2 \right)\,,
\end{align}
 and normalisation
\begin{align}
    \int d^4 x \sqrt{\bar{g}} {h^{(n)}_{\mu \nu}}^* {\bar{g}}^{\mu \alpha}{\bar{g}}^{\nu \beta} h^{(m)}_{\alpha \beta} = \delta^{mn} \,.
\end{align}
These are modes that annihilate $D$, while still being normalisable and smooth. For these reasons, $h^{(n)}_{\mu \nu}$ are zero modes of the Euclidean path integral of the background solution \eqref{EqdSgeometry}. 

Finally, we can quantify how the eigenvalues of these modes are affected by the temperature deformation in \eqref{eq:RN-Nariai-near} via \eqref{eq:correc-D}. As in \eqref{EqSdShDhCoinc}, we find a negative sign in the finite temperature contribution to the graviton operator
\begin{align}
  \int d^4x \sqrt{g} {h^{(n)}_{\alpha \beta}}^* \delta D^{\alpha \beta \mu \nu} h^{(n)}_{\mu \nu} = - \frac{|n| T_c}{16r_N} + \mathcal{O} \left( T_c^2 \right) \,.
\label{EqdShDh}
\end{align}
Therefore once again, the tensor modes introduce a non-convergent Gaussian contribution to the path integral as $(\delta\Lambda_{n})_{\text{tensor}}<0$.
\paragraph{Vector modes.}
\label{SecdSVector}
To construct the vector modes we now solve \eqref{EqLargeGauge} on \eqref{EqdSgeometry}, obtaining
\begin{align}
    \Phi^0_n = \frac{1}{\sqrt{2\pi^3 |n|} r_N^2} \left( \frac{\sinh \eta}{1+\cosh \eta} \right)^n e^{in\tau} \,,
\label{EqNariaiGaugeScalar}
\end{align}
where as before $\left| n \right| \geq 1$. The resulting vector modes $v_{\mu \nu}^{(m,n)}$ are simply given by \eqref{eq:vector-mode}. As in \eqref{EqSdSvecnorm}, the norm of these modes is negative: 
\begin{align}
    \int d^4x \sqrt{\bar{g}} {v_{\mu \nu}^{(m,n)}}^* \bar{g}^{\mu \alpha} \bar{g}^{\nu \beta} v_{\alpha \beta}^{(m',n')} = - \delta^{nn'}\delta^{mm'} \,.
\label{EqdSVectornorms}
\end{align}
Still these modes annihilate $D$ in \eqref{EqLinGravOperator} and are orthogonal to tensor modes $h^{(n)}_{\mu \nu}$.

Despite the negativity in \eqref{EqdSVectornorms}, we can study their response under the temperature deformation \eqref{eq:RN-Nariai-near} by evaluating \eqref{eq:correc-D}. Similar to prior cases, we find that for all three values of $m$, these modes make the same finite temperature contribution 
\begin{align}
    \int d^4x \sqrt{g} {v^{(m,n)}_{\alpha \beta}}^* \delta D^{\alpha \beta \mu \nu} v^{(m,n)}_{\mu \nu} = \frac{\left( 2r_N^2 - \ell_{\text{dS}}^2 \right) |n|T_c}{48r_N^3} + \mathcal{O} \left( T_c^2 \right) \,.
\label{EqdSVectorvDv}
\end{align}
The vector mode contribution has a change in sign, depending on which side of the lukewarm curve the black hole sits.\footnote{Recall that $Q_N = M_N$ corresponds to $2r_N^2 - \ell_{\text{dS}}^2=0$, and we have $\left( T_+ - T_c \right)  \propto \left( 2r_N^2 -\ell_{\text{dS}}^2 \right) T_c^2+\cdots$.} This change in sign is represented by the $\pm$ in Table \ref{fig:table1}. For $T_+ > T_c$, we have a positive numerator in \eqref{EqdSVectorvDv}, and for $T_+ < T_c$ it is negative. We note then that at the lukewarm point, $Q_N = M_N$, the vector mode contribution vanishes to leading order in the temperature.

\paragraph{Gauge (photon) modes.}
\label{SecdSPhoton}
An important difference between the RNdS$_4$ and SdS$_4$ cases is that we can construct photon modes. Analogous to the Cold limit, we construct our photon modes from the scalars \eqref{EqNariaiGaugeScalar} generating large gauge transformations as
\begin{align}
    {a^{(n)}} = \frac{\sqrt{\pi} r_N}{2} \partial_{a}\Phi_n^0 \left( \tau, \eta \right) \d x^a \,.
\label{EqdSamodes}
\end{align}
These modes are normalised such that
\begin{align}
    \int d^4 x \sqrt{\bar{g}} {a^{(n)}_{\mu}}^* {\bar{g}}^{\mu \nu} {a^{(n')}_{\nu}} = -\delta^{nn'} \,,
\label{EqdSauNorms}
\end{align}
where the norm is \textit{negative} and as before $\left| n \right| \geq 1$. We can see from equation \eqref{EqdSamodes} that the photon modes are defined only in terms of the $-\text{EAdS}_2$ part of the metric, and in the norm in \eqref{EqdSauNorms} a minus sign arises. By comparison, whilst the tensor modes were also only defined in the $-\text{EAdS}_2$ part of the metric, those being two-component tensors means two minus signs cancelled out. 

As in Sec.\,\ref{sec:coldphoton}, the $a^{(n)}$ are subleading in $Q_N$ relative to the background field $\bar{A}$ at the boundary $\eta \rightarrow \infty$. Similarly, they annihilate the operator $P$ in \eqref{EqPhotonOperator}, and $O_{\text{int}}$ in \eqref{EqIntha} at extremality ($T_c=0$), and so they are zero modes of the path integral.

We compute the contribution of these modes to the photon operator to be 
\begin{align}
    \int d^4 x \sqrt{g} {a_{\mu}^{(n)}} \delta P^{\mu \nu} a_{\nu}^{(n)} = - \frac{\left( r_N^2 + \ell_{\text{dS}}^2 \right) |n| T_c}{24r_N^3} + \mathcal{O} \left( T_c^2 \right) \, .
\label{EqaDaNariai}
\end{align}
It is interesting to contrast \eqref{EqaDaNariai} with the correction to the vector mode \eqref{EqdSVectorvDv}. The lukewarm point is not special here. Also, the correction to the eigenvalue $(\delta\Lambda_{n})_{
\text{gauge}}$ is positive in \eqref{EqaDaNariai}. This means that for the gauge modes, we could restore the convergence of the path integral by changing the definition of the norm in \eqref{EqdSauNorms} as was done in, for example, \cite{Kolanowski:2024zrq} where a similar issue arose.

As a final remark, it is worth noting that the expressions in \eqref{EqdShDh}, \eqref{EqdSVectorvDv}, and \eqref{EqaDaNariai}, under the map \eqref{eq:map-cold-N}, are compatible with their cold counterparts \eqref{EqAdShDh}, \eqref{EqAdSvDv}, and \eqref{EqaDaCold}, provided one accounts for the norms $\kappa_n$ and $\gamma_n$ (see \eqref{eq:correc-D} and \eqref{eq:correc-P}). That is, schematically, under the map \eqref{eq:map-cold-N}, we find
\begin{equation}
    \ell_{A}^2 \rightarrow -\ell_{\text{dS}}^2\,,\quad r_0 \rightarrow r_N\,,\quad T_+ \rightarrow - T_c \,,
    \quad
    \Rightarrow
    \quad
    \left( \delta \Lambda \right)_{\text{Cold}} \rightarrow \left( \delta \Lambda \right)_{\text{Nariai}} \, ,
\end{equation}
where we comment that the map also works in the reverse direction.

\subsection{Complexified near-Nariai background: \texorpdfstring{$-(\text{EAdS}_2 \times S^2)$}{}}
\label{SubSecWickRot}

The essence of our dilemma is the signature of the Euclidean geometry in \eqref{EqNariaiGeometryEuclidean1}, which is $(-,-,+,+)$; this percolates into various corners of the computation in Sec.\,\ref{sec:decoup61} which renders negative eigenvalues and norms.  The way we will address this is by an analytic continuation of the geometry that lands us on \eqref{EqMinusEAdS2S2}, which has now signature $(-,-,-,-)$. This is what we denote as $-(\text{EAdS}_2 \times S^2)$. This background effectively rotates the contour of integration in our path integral, and we will see that the integral over certain zero modes will be convergent in an appropriate way.\footnote{Another standard option is to analytically continue answers from AdS to dS \cite{Maldacena:2002vr}. We are purposely not taking this route since we would like to resolve this problem while staying in de Sitter.} 

In this first portion, we will describe how to treat the RNdS$_4$ black hole such that we obtain a heated version of \eqref{EqMinusEAdS2S2} and incorporate the appropriate deformations.  The simplest way to do so is to take $r_N\to i \rho_N$ in \eqref{geom_Nariai_exp}-\eqref{eq:RN-Nariai-near}.\footnote{One can define a suitable decoupling limit starting for the RNdS$_4$ leading to the same answer \eqref{geom_Nariai_exp_wick}-\eqref{eq:wick-N-deformed}. The idea is to take $r\to i \rho$, $r_+\to i\rho_+$ and $r_c\to i \rho_c$, while keeping $\ell_4$ and $t$ unchanged, in \eqref{eq:4dvanillametric}-\eqref{eq:gaugefield}. There is then a decoupling limit for $\rho$ that has a similar form as in  \eqref{eq:newcoords1}. We omit it here for brevity.} The resulting metric and gauge field are
\begin{equation}
    \begin{aligned} \label{geom_Nariai_exp_wick}
    g_{\mu \nu} & = \bar{g}_{\mu \nu} + \tilde{T_c}\,\delta \tilde{g}_{\mu \nu}  \,   +  \mathcal{O} \left( \tilde T_c^2 \right)\,,  \\
    A &= \bar{A} + \tilde{T_c} \,\delta \tilde{A} \,  + \mathcal{O} \left( \tilde T_c^2 \right)\,,
\end{aligned}
\end{equation}
where
\begin{equation}
    \begin{aligned}
   \bar{g}_{\mu\nu} \d x^\mu \d x^\nu &=  - \ell_{\text{dS}}^2 \left( \sinh^2 \eta \d \tau^2 + \d \eta^2 \right)  - \rho_N^2 d\Omega_2^2 \,, \label{EqdSgeometry_wick} \\
 \bar A &=  - i Q_N \frac{ \ell_{\text{dS}}^2}{\rho_N^2} \left( \cosh \eta -1 \right) \, \d\tau\,.
\end{aligned}
\end{equation}
We have also defined
\begin{equation}
    \tilde T_c= i T_c~,
\end{equation}
and the effective length $\ell_{\text{dS}}$ is now given by
\begin{align}
    \ell_{\text{dS}}^2 = \frac{\ell_4^2 \rho_N^2}{ 6\rho _N^2 + \ell_4^2} \, .
\label{Eqldsrho}
\end{align}
As promised, we have a $-(\text{EAdS}_2\times S^2)$ near horizon geometry at extremality, where the relative minus sign between the two parts of the metric has been removed. Now $\tilde T_c$ takes the role of temperature, where the leading order response  reads
\begin{equation}
    \begin{aligned}\label{eq:wick-N-deformed}
  \frac{ \delta \tilde{g}_{\mu\nu} \d x^\mu \d x^\nu}{4\pi \ell_{\text{dS}}^2} &=    \frac{\ell_{\text{dS}}^2  }{3\rho_N^3} \left( \ell_{\text{dS}}^2+2\rho _N^2 \right)\left( 2 + \cosh \eta \right)\tanh^2 \frac{\eta}{2}\left( - \sinh^2 \eta \d \tau^2 + \d \eta^2  \right) \\&\qquad + \rho _N \cosh \eta d\Omega_2^2 \,,\\
 \delta \tilde{A} &=  - 2\pi i \frac{ \ell_{\text{dS}}^4}{\rho_N^3} Q_N \sinh^2 \eta \,d\tau\,,
\end{aligned}
\end{equation}
After this Wick rotation, notice that $\delta g_{\tau\tau}$ and $\delta g_{\eta\eta}$ have a definite sign; our complexification of the coordinates has removed the lukewarm point which appears in \eqref{eq:RN-Nariai-near}. 
Also, the electric charge now takes the value
\begin{align}
    Q_N^2 = - \frac{\rho_N^2}{2\ell_{\text{dS}}^2} \left( \ell_{\text{dS}}^2 + \rho_N^2  \right) \, ,
\label{EqQNrho}
\end{align}
hence for real $\rho_N$, $Q_N$ and $M_N$ are imaginary. This indicates that we are now studying a complex black hole geometry.

Mathematically, the $-(\text{EAdS}_2\times S^2)$ geometry in equation \eqref{geom_Nariai_exp_wick} is a Euclidean solution to Einstein's equations. If we wish to consider $\rho$ and $\tilde{T}_c$ as real parameters (that is, $T_c$ is purely \textit{imaginary}), then $\rho_+$ and $\rho_c$ are purely real as well. This means that the horizons $r_+$ and $r_c$ have been deformed into the complex plane, and we are considering some complexified black hole geometry. Note that $\rho_+ > \rho_c$, however in the instance that $\tilde{T}_c $ is real these no longer correspond to physical horizons, so it is unclear if the change in order is important.

\paragraph{Tensor modes.} Firstly, for the tensor modes, the norm remains positive, but the graviton operator contribution now has the sign which leads to a Gaussian integral. In particular, the tensor modes become
\begin{align}
    h^{(n)}_{\mu \nu} \d x^{\mu} \d x^{\nu} = & \frac{e^{in\tau}\ell_{\text{dS}} \sqrt{|n|(n^2-1)}}{\sqrt{8}\pi i \rho_N} \tanh^n \frac{\eta}{2} \left( \frac{\d \eta^2}{\sinh^2 \eta} + \frac{2i\d \eta \d \tau}{\sinh \eta} - \d \tau^2 \right),
\end{align}
with $\left| n \right| \geq 2$ and their norm is still well-behaved
\begin{align}
    \int d^4 x \sqrt{\bar{g}} {h^{(n)}_{\mu \nu}}^* {\bar{g}}^{\mu \alpha}{\bar{g}}^{\nu \beta} h^{(n')}_{\alpha \beta} = \delta^{n n'}.
\end{align}
What is significant is that the correction to the eigenvalue becomes positive
\begin{align}
    \begin{split}
        \left( \delta \Lambda_n \right)_{\text{graviton}} = \int d^4x \sqrt{g} {h^{(n)}_{\alpha \beta}}^* \delta D^{\alpha \beta \mu \nu} h^{(n)}_{\mu \nu} = & \frac{|n| \tilde{T}_c}{16 \rho_N}   + \mathcal{O} \left( \tilde{T}_c^2 \right) .
    \end{split}
\label{EqdShDhWick}
\end{align}
Choosing $\tilde{T}_c $ real and $\tilde{T}_c > 0$ means we have a convergent path integral. Computing the path integral and using the same zeta function regularisation as in Sec.\,\ref{SecColdLimit}, we obtain
\begin{align}
    \left( \delta \log Z \right)_{\text{tensor}} = & \frac{3}{2} \log \frac{ \tilde{T}_c}{\rho_N } - 3 \log \left( 4 \sqrt{6\pi} \right) + \dots \, .
    \label{EqCompTens}
\end{align}

\paragraph{Vector modes.} The Wick rotation also resolves the issue with the norm of the vector modes. On our new background, the scalar field solving \eqref{EqLargeGauge} becomes
\begin{align}
    \Phi^{0}_n = - \frac{1}{\sqrt{2\pi^3 |n|} \rho_N^2} \left( \frac{\sinh \eta}{1+\cosh \eta} \right)^n e^{in\tau} \, ,
\end{align}
where $\left| n \right| \geq 1$ and the norm of the modes is now \textit{positive}
\begin{align}
    \int d^4x \sqrt{\bar{g}} {v_{\mu \nu}^{(m,n)}}^* \bar{g}^{\mu \alpha} \bar{g}^{\nu \beta} v_{\alpha \beta}^{(m',n')} = \delta^{m m'} \delta^{n n'} \,.
\end{align}
We proceed as before and compute the leading order in $\tilde{T}_c$ contribution to the graviton operator to be
\begin{align}
    \begin{split}
        \left( \delta \Lambda_n \right)_{\text{vector}} = \int d^4x \sqrt{g} {v^{(m,n)}_{\alpha \beta}}^* \delta D^{\alpha \beta \mu \nu} v^{(m,n)}_{\mu \nu}  = & \frac{\left( 2\rho_N^2 + \ell_{\text{dS}}^2 \right) }{48 \rho_N^3} |n| \tilde{T}_c + \mathcal{O} \left( \tilde{T}_c^2 \right) \\
        = & \frac{4 \rho_N^2 + \ell_4^2}{6\rho_N^2 + \ell_{4}^2} \frac{|n|  \tilde{T}_c}{16\rho_N}   + \mathcal{O} \left( \tilde{T}_c^2 \right)~.
    \end{split}
\label{EqdSvDvWick}
\end{align}
In the last line, we have used \eqref{Eqldsrho} to make it explicit that equations \eqref{EqdShDhWick} and \eqref{EqdSvDvWick} now have the same sign. Note that, due to the removal of the lukewarm point, this contribution now has a \textit{definite} sign, as opposed to the three different regimes in \eqref{EqdSVectorvDv}. Computing the convergent path integral, we report
\begin{align}
    \left( \delta \log Z \right)_{\text{vector,m}} = & \frac{1}{2} \log \frac{\left( 2 \rho_N^2 + \ell_{\text{dS}}^2 \right) \tilde{T}_c }{\rho_N^3} - \log \left( 4 \sqrt{3 \pi} \right) + \dots \, .
    \label{EqCompVec}
\end{align}

\paragraph{Gauge (photon) modes.} The story is not quite as neat for the photon modes. In particular, the modes
\begin{align}
    {a^{(n)}} = \frac{i \sqrt{\pi} \rho_N}{2} \partial_{a}\Phi_n \left( \tau, \eta \right) \d x^a \,,
\label{EqdSamodes_wick}
\end{align}
where $\left| n \right| \geq 1$ still have negative norm
\begin{align}
    \int d^4x \sqrt{\bar{g}} {a^{(n)}_{\mu}}^* {\bar{g}}^{\mu \nu} {a^{(n')}_{\nu}} = - \delta^{n n'}.
\label{EqPhotonWickNorm}
\end{align}
However, the leading order correction 
\begin{align}
    \begin{split}
        \int d^4 x \sqrt{g} {a_{\mu}^{(n)}}^* \delta P^{\mu \nu} a_{\nu}^{(n)} = & \frac{\left( \rho_N^2 - \ell_{\text{dS}}^2 \right)}{24 \rho_N^3 }  |n|  \tilde{T}_c  + \mathcal{O} \left( \tilde{T}_c^2 \right) \\
        = & \frac{4\rho_N^2}{6\rho_N^2 + \ell_4^2} \frac{|n| \tilde{T}_c }{16 \rho_N }  + \mathcal{O} \left( \tilde{T}_c ^2 \right)\,,
    \end{split}
\label{EqdSaDaWick}
\end{align}
which would lead to a \textit{negative} eigenvalue via \eqref{eq:correc-P}.
In the final line, we have used \eqref{Eqldsrho} to make explicit that the contribution \eqref{EqdSaDaWick} has a definite sign. 

\paragraph{Considering all zero mode contributions.} Firstly, let us take a moment to understand what has happened here. By Wick rotating $\rho_N$, we have introduced a factor of $(-1)$ when integrating over the $S_2$. This introduces a factor of $(-1)$ to the norm of the vector modes, making that norm positive. The Wick rotation also introduces a factor of $(-1)$ to each operator, which flips the sign of the tensor and gauge modes. The two factors of $(-1)^2 = 1$ preserve the positive sign of the vector mode contribution. 

We noted that the Wick rotation does not resolve the issue of the photon modes having minus sign. One therefore might be motivated to consider the $Q \rightarrow 0$ limit of the theory, such that there are no gauge fields with negative norm. This is of course the SdS$_4$ geometry we discussed in Sec.\,\ref{SecSdSNariai} --- so we should indeed recover the problems we found in that case. In the $Q = 0$ limit of our Wick rotated geometry, $\ell_4^2 = 3 r_N^2 = - 3 \rho_N^2$ and $\ell_{\text{dS}}^2 = - \rho_N^2$. Therefore, the metric rotates to EAdS$_2 \times (-S^2)$, and the problematic relative minus sign difference reappears (just switched between the two halves of the metric). As a consequence, the norm of the vector modes once again becomes minus one (c.f. equations \eqref{EqSdSvecnorm} and \eqref{EqdSVectornorms}). Additionally, the tensor and vector mode contributions are now
\begin{equation}
    \begin{aligned}
        \int d^4 x \sqrt{g} {h^{(n)}_{\alpha \beta}}^* \delta D^{\alpha \beta \mu \nu} h^{(n)}_{\mu \nu}  = & \frac{ |n|  \tilde{T}_c }{16  \rho_N}   + \mathcal{O} \left( \tilde{T}_c^2 \right), \\
        \int d^4 x \sqrt{g} {v^{(m,n)}_{\alpha \beta}}^* \delta D^{\alpha \beta \mu \nu} v^{(m,n)}_{\mu \nu}  = & -\frac{|n| \tilde{T}_c}{48 \rho_N}   + \mathcal{O} \left( \tilde{T}_c^2 \right).
    \end{aligned}
\end{equation}
That is, from \eqref{eq:correc-D}, we have a positive eigenvalue correction for both the tensor and vector modes. If we change the definition of the inner product of $v^{(m,n)}$ along the lines of  \cite{Marolf:2022ntb,Liu:2023jvm,Kolanowski:2024zrq}, we will therefore obtain, in the $Q = 0$ limit of the Wick rotated geometry, a convergent path integral.

It is worth noting that in two-dimensional JT gravity, only the tensor modes are visible (such as in \cite{Maldacena:2019cbz,Cotler:2019nbi}), this problem would be hidden and one could choose $\tilde{T}_c$ and $\rho_N$ to have appropriate sign such that the 2D path integral converged.  However, we are interested in a convergent path integral for the 4D theory that includes vector and gauge modes. These can be added as additional massless modes, however we are encountering that we don't have enough free parameters in the effective action for these modes to render each term convergent. It seems like the best case scenario is to not include the modes  $a^{(n)}$ in the path integral.

\section{Conclusions}\label{sec:conclusion}

In this work, we have computed quantum corrections to the gravitational partition function for the near horizon geometries of various RNdS$_4$ configurations near extremality. 
These quantum corrections originate from zero modes, associated to diffeomorphism and gauge symmetries that persist after gauge fixing. 
In the setup we adopt, we are able to consider contributions from tensor modes, vector modes and gauge modes. 
We summarize our findings in the various limits in Table \ref{fig:table2}.

\begin{table}[ht]
\begin{center}
 \begin{tabular}{|c|c|c|}
\hline
Decoupling limits                                                     & \begin{tabular}[c]{@{}c@{}}Near-horizon\\ geometry\end{tabular} & \begin{tabular}[c]{@{}c@{}}Partition function\\ $Z_{\text{low-$T$}}\propto$\end{tabular} \\ \hline \hline
RNdS$_4$ Cold                                                            & EAdS$_{2}\times S^2$                                                          & $T^{7/2}$                                                             \\ \hline
SdS$_4$ Nariai                                                           & $-\text{EAdS}_{2}\times S^2$                                                           & unresolved  
 \\ \hline
\begin{tabular}[c]{@{}c@{}}SdS$_4$ Nariai\\ (Complexified)\end{tabular} & $\text{EAdS}_{2}\times(- S^2)$                                                           & $T^3$    
\\ \hline
RNdS$_4$ Nariai                                                          & $(-\text{EAdS}_{2})\times S^2$                                                           & unresolved                                                         \\ \hline
\begin{tabular}[c]{@{}c@{}}RNdS$_4$ Nariai\\ (Complexified)\end{tabular} & $-(\text{EAdS}_{2}\times S^2)$                                                           & unresolved                                                              \\ \hline
\end{tabular}

\end{center}
\caption{In this table we complement the details presented in Table \ref{fig:table1} with the near-horizon geometry and the resulting partition function.}
\label{fig:table2}
\end{table}

In the Cold limit, where the inner and outer horizon of the black hole are coincident, we found a contribution from the gauge modes that is not present for the usual Reissner-Norström case in asymptotically flat spacetime. This resulted in a $T^{7/2}$ correction to the partition function at low temperatures, unlike the $T^{3}$ correction reported in \cite{Banerjee:2023quv} for asymptotically flat spacetimes. We also verified that this contribution persists for a Reissner-Nordström black hole in AdS$_4$ spacetime.
We note that our machinery depends on zooming in on the near-horizon region. 
In the Cold case, this procedure allowed us to ignore dealing with the cosmological horizon, as it is not contained in the emerging AdS$_{2}\times S^{2}$ picture. That is, by zooming in, we circumvent potentially thorny issues of trying to deal with an Euclidean geometry that cannot be smooth at both the outer black hole horizon and the cosmological horizon. There are recent developments on how to define thermodynamics with multiple horizons in de Sitter. For example, one could introduce an observer on a timelike boundary \cite{Banihashemi:2022jys}, or consider an instanton formulation of the path integral \cite{Morvan:2022aon,Morvan:2022ybp}.  It would be interesting to explore this so we can drop the crutch of zooming in on the near-horizon region to study the robustness of the results presented here. Zooming out of an AdS$_2$ throat was explored in asymptotically Anti-de Sitter and flat spacetimes in \cite{Kapec:2024zdj,Kolanowski:2024zrq,Arnaudo:2024bbd}; see also \cite{Turiaci:2025xwi} where this is applied to the wave function of an $S^1\times S^2$ universe in dS$_4$.

We then turned our attention to the Nariai limit.
There, the cosmological and outer black hole horizons are coincident. 
We worked outside the cosmological horizon, i.e., the Milne patch, because the Euclidean static patch is a compact geometry (up to a conical defect) which should not contain zero modes. 
When working on $(-\text{EAdS}_{2})\times S^2$ we found, however, that both the charged and uncharged cases had two pathologies: we have modes with negative eigenvalues and modes with a negative norm. A simple resolution is to just deform the contour of integration and modify definitions such that the path integral is convergent. In that case, we would report that for RNdS$_4$ in the Nariai limit, we have $\log Z\sim 7/2 \log T$, and for SdS$_4$ in the Nariai limit, we have $\log Z\sim 3 \log T$.   However, we prefer to leave this ``unresolved,''  as displayed in Table \ref{fig:table2}, since de Sitter quantum gravity is delicate and we do not have a microscopic model to contrast this with. It would be interesting to understand our computations from the perspective of the static patch: if the log-$T$ contribution is physical one should be able to capture it by deforming $S^2\times S^2$ and this would reassure that the appropriate adjustment to the path integral in the Milne patch are sound. One way to address the log-$T$ contribution in the static patch might be via the methods of quasinormal modes used in \cite{Kapec:2024zdj,Arnaudo:2024bbd}. We did find that we could remedy some aspects of the ill-defined zero mode contributions and the negative norm of the vector modes by complexifying the geometry. In the case of SdS$_4$ this provides a cure and hence a reasonable way to deform the contour of integration of the path integral. Nevertheless, in the RNdS$_4$ we will have issues with the gauge modes: on the complex geometry, they have both negative eigenvalues and a negative norm.

One last observation is in order here: the Nariai configuration is characterized by {\it negative} specific heat, as one can see from the low-temperature expansion of the thermodynamic quantities and in \eqref{thermo-rels-Nariai1}. In this case, (nonconformal) modes with a negative contribution to the gravitational path integral are expected \cite{Prestidge:1999uq,Gross:1982cv}. This is in addition to the fact that generically the conformal mode of the metric (trace) has the wrong sign in the kinetic term, which usually is circumvented by integrating the conformal mode along a purely imaginary direction. The transverse traceless modes in the most usual ensembles couple to the conformal mode and create problems when one wants to integrate along the purely imaginary direction but further studies \cite{Marolf:2022ntb,Liu:2023jvm} resolved this issue (see also discussion in \cite{Kolanowski:2024zrq}). We expect this to be the case also in our analysis, even though this point deserves a deeper study.

 As noted in Sec.\,\ref{SecNariaiCorrections}, the Nariai limit of the SdS$_4$ geometry has been studied in \cite{Maldacena:2019cbz,Cotler:2019nbi,Moitra:2022glw} using the boundary Schwarzian theory to compute the \textit{Lorentzian} path integral. Physically, this has an interpretation as a wavefunction in Milne patch, which \cite{Maldacena:2019cbz} note could be identified as a solution to a two-dimensional Wheeler-DeWitt (WDW) equation. Whilst our machinery is for a Euclidean path integral, and thus naturally suited to computing partition functions, in minisuperspace it is possible to interpret static patch partition functions as the analytic extension of Milne patch wavefunctions via analytic continuation through the cosmological horizon \cite{Blacker:2023oan}. It would be interesting to construct a possible WDW interpretation of our computations, which may be helpful for resolving some of the complications uncovered.

In this work, we reported on quantum corrections to the path integral in the Cold and Nariai limits. We are yet to discuss here the quantum corrections in the Ultracold limit introduced in Sec.\,\ref{SecUCGeometry}, where all three horizons coincide. To consider quantum corrections to the geometry there, it is necessary to work in a grand canonical ensemble (with fixed chemical potential $\mu$), as any change of the temperature at fixed charge $Q$ takes us out of the allowed region of the Shark Fin in Fig.\,\ref{fig:sharkfin}. When attempting to construct the relevant zero modes, two issues arise. Firstly, the transverse-traceless gauge in \eqref{EqGuageTT} imposes too many constraints on the graviton perturbations, and it is not possible to construct diffeomorphisms that would mimic the BMS$_2$ group associated with $\text{Mink}_2$. This problem arises irrespective of how we describe the background metric $\bar g_{ab}$. A similar problem arose in \cite{Bagchi:2012yk,Detournay:2014fva} for $\text{Mink}_3$, and in fact it is simply a coincidence that in $\text{(A)dS}_2$ one can recover the asymptotic symmetries in transverse-traceless gauge\footnote{We thank Daniel Grumiller for advice on this particular point}. 
Secondly, the diffeomorphisms and gauge transformations that are compatible with \eqref{EqGuageTT} are non-normalisable. Specifically, their norm contains an IR or UV divergence.\footnote{In this analysis we did not impose an ensemble to fix boundary conditions. The analysis was done in a fairly general way.} We are presently working on resolving these issues and computing the full set of quantum corrections in the $\text{Mink}_2 \times S^2$ geometry. Logarithmic corrections have been predicted from a boundary perspective (see for example \cite{Afshar:2021qvi}), which should provide a lamppost in our efforts. We expect that one needs to modify \eqref{EqGuageTT} and re-work the appropriate one-loop effective action.


\acknowledgments
We thank G. Bonelli, R. Emparan, D. Grumiller, M. Heydeman, L. Iliesiu,  S. Murthy, M. Rangamani, M. A. Tanzini, M. Tomasevic, G. Turiaci, and Z. Yang for interesting discussions and collaborations on related topics. The work of CT is supported by the Marie Sklodowska-Curie Global Fellowship (REA Horizon 2020 Program) SPINBH MICRO-101024314. This work was done in part during the Aspen workshop on ``Microscopic origin of black hole entropy", Aspen, CO  (2024) and the workshop ``Holographic Duality and Models of Quantum Computation" held at Tsinghua Southeast Asia Center in Bali, Indonesia (2024). The work of MJB is supported by a Gates Cambridge Scholarship (OPP1144). The work of WS is supported by a Starting Grant 2023-03373 from the Swedish Research Council.  MJB, AC and WS have been partially supported by STFC consolidated grant ST/X000664/1.

 \begin{appendix}

\section{Ghosts}\label{app:ghost}

In this appendix, we argue why the ghost Lagrangian \eqref{EqLagGhost}, for backgrounds of the form \eqref{eq:background}, does not give rise to zero modes.
The ghost Lagrangian has the same form as the one in flat space, and it reads \cite{Banerjee:2011jp,Sen:2012kpz}
\begin{equation}
    \label{ghost_act}
\mathcal{L}_{\text{ghost}} = b^{\mu} (\bar{g}_{\mu \nu} \Box + R_{\mu\nu}) c^{\nu}  + b \Box c -2b \bar{F}^{\mu \nu} \nabla_{\mu} c_{\nu} \, ,
\end{equation}
where the ghost fields $\lbrace b_{\mu}, c_{\mu} \rbrace$ are associated with diffeomorphism invariance and $\lbrace b, c \rbrace$ are associated with $U(1)$ gauge invariance. Following the reasoning in \cite{Banerjee:2011jp}, the last term in \eqref{ghost_act} gives mixing between the fields $b$ and $c^{\mu}$, but the mixing matrix has an upper triangular form, hence this term gives no contribution. Additionally, $b \Box c$ cannot give a zero mode contribution as the photon ghosts are scalars. Turning to the graviton ghosts, we can expand them in the basis \cite{Banerjee:2011jp}
\begin{equation}
    \begin{aligned}
        \label{ghost_mode}
b_{i} = & \frac{A}{\sqrt{\kappa_1}} \, \partial_{i} u + \frac{B}{\sqrt{\kappa_1}}  \, \epsilon_{ij} \partial^{j} u \, , \\
b_{a} = & \frac{C}{\sqrt{\kappa_2}} \partial_{a} u + \frac{D}{\sqrt{\kappa_2}}  \, \epsilon_{ab} \partial^{b} u  \, , \\
c_{i} = & \frac{E}{\sqrt{\kappa_1}} \partial_{i} u + \frac{F}{\sqrt{\kappa_1}}  \, \epsilon_{ij} \partial^{j} u \, ,\\
c_{a} = & \frac{G}{\sqrt{\kappa_2}} \partial_{a} u + \frac{H}{\sqrt{\kappa_2}}  \, \epsilon_{ab} \partial^{b} u  \, ,
    \end{aligned}
\end{equation}
where we indicated with $i,j$ the coordinates on $S^2$ and with $a,b$ those on the $AdS_2$ manifold. The field $u$ is the product 
\begin{equation}
    u = f_{\lambda, l} (\tau, \eta) Y_{l,m} (\theta, \phi) \, ,
\end{equation}
of the spherical harmonics $Y_{lm}$ and the $AdS_2$ Laplacian eigenfunctions $f_{\lambda l}$ that satisfy respectively
\begin{equation}\label{various_box}
\Box Y_{lm} = - \kappa_1 Y_{l,m} \, ,\qquad \Box f_{\lambda, l} = -\kappa_2 f_{\lambda, l} \, ,
\end{equation}
with
\begin{equation}
\label{values_kappas}
\kappa_1 =\frac{l (l+1)}{ r_0^2} \, ,  \qquad  \kappa_2 =  \left( \lambda^2 + \frac14 \right) \frac{1}{\ell_{\text{AdS}}^2}\, .
\end{equation}
In \eqref{ghost_act} the 4d Laplacian comes from the metric
\begin{equation}
    ds^2 = \bar{g}_{\mu \nu} \d x^{\nu} \d x^{\nu} = \ell_{\text{AdS}}^2 (\sinh^2\eta \, \d \tau^2 + \d \eta^2) + r_0^2 (\d \psi^2 + \sin^2\psi \d \phi^2) \, ,
\end{equation}
and substituting the modes \eqref{ghost_mode} into the ghost action \eqref{ghost_act}, repackaging the various parts in term of the 2d Laplacians, using \eqref{various_box} and focussing only on the terms ABEF we get a contribution
\begin{equation}
    \label{res_ghost}
\left( \kappa_2+ \kappa_1 - \frac{2}{r_0^2} \right) (AE+BF) \,,
\end{equation}
from which one recovers the flat space result in \cite{Banerjee:2011jp} 
for $\ell_{\text{AdS}} = r_0$. The terms CDGH instead give a contribution
\begin{equation}
    \label{res_ghost2}
\left( \kappa_2+ \kappa_1 + \frac{2}{\ell_{\text{AdS}}^2} \right) (CG+DH)\, ,
\end{equation}
which we verified for $l=0$. Being a sum of positive terms, the contribution \eqref{res_ghost2} does not give rise to zero modes.

We need then to analyze \eqref{res_ghost}. Notice that for this term we need to have $l \geq 1$ (the $l=0$ modes do not exist). By taking into account \eqref{values_kappas} 
we get that the value of this first term in \eqref{res_ghost} is always positive and strictly larger than zero since $\lambda$ is real. So we can conclude that the modes \eqref{ghost_mode} cannot give rise in this context to additional zero modes.

 \end{appendix}
\bibliographystyle{JHEP}
\bibliography{new-bib}

\end{document}